\documentclass[iop,natbib209]{emulateapj}

\usepackage[usenames,dvipsnames,svgnames,table]{xcolor}

\usepackage{graphicx}
\usepackage{amsmath}
\usepackage{natbib}

\shorttitle{}
\shortauthors{Hsieh et al.}

\begin{document}

\title{A Magnetic Field Connecting the Galactic Center Circumnuclear Disk with Streamers and Mini-spiral -
Implications from 850 $\micron$ Polarization Data}

\author{
        Pei-Ying Hsieh\altaffilmark{1},
        Patrick M. Koch\altaffilmark{1},
        Woong-Tae Kim \altaffilmark{2,3},
        Paul T. P. Ho\altaffilmark{1,4},
        Ya-Wen Tang\altaffilmark{1},
        Hsiang-Hsu Wang \altaffilmark{5}
\\pyhsieh@asiaa.sinica.edu.tw}

\affil{$^1$ Academia Sinica Institute of Astronomy and
       Astrophysics, P.O. Box 23-141, Taipei 10617, Taiwan, R.O.C.}
\affil{$^2$ Department of Physics \& Astronomy, Seoul National University, Seoul 151-742, Korea}
\affil{$^3$ Department of Astrophysical Sciences, Princeton University, Princeton, NJ 08544, USA}
\affil{$^4$ East Asian Observatory, 660 N. Aohoku Place, University Park, Hilo, Hawaii 96720, U.S.A.}
\affil{$^5$ Department of Physics and Institute of Theoretical Physics, The Chinese University of Hong Kong, Shatin, New Territories, Hong Kong, China}

\begin{abstract}

Utilizing James Clark Maxwell Telescope (JCMT) 850 $\micron$ SCUPOL dust polarization data, we investigate the configuration of the magnetic ($B$) field in the circumnuclear disk (CND) of the  Galactic Center (GC). The SCUPOL data show a highly improved polarization coverage and resolution compared to earlier 100 $\micron$ observations. The 850 $\micron$ data have a resolution and coverage similar to previous 350 $\micron$ polarimetry data. However, with a proper sampling on a 10$"$ grid, we find the 850 $\micron$ data trace the morphological structures of the CND substantially better. Furthermore, because the 850 $\micron$ trace the field deeper into the material near Sgr A*, they represent the highest resolution submillimeter probe to date of the CND magnetic field. The observed $B$-field morphology is well described by a self-similar axisymmetric disk model where the radial infall velocity is one quarter of the rotational velocity. A detailed comparison with higher-resolution interferometric maps from the Submillimeter Array further reveals that the $B$-field aligns with the neutral gas streamers connecting to the CND.  Moreover, the innermost observed $B$-field structure also appears to trace and align with the mini-spiral located inside the CND.  This suggests that there is one underlying $B$-field structure that is connecting the CND with its streamers and the inner mini-spiral. An estimate of $\beta_{\rm Plasma} \lesssim 1$ -- based on the global B-field morphology that constrains  the azimuthal-to-vertical field strength ratio of around 40 combined with a measurement of the azimuthal velocity -- indicates that the B-field appears dynamically significant towards the CND and also onwards to the inner mini-spiral.

\end{abstract}

\keywords{Galaxy: center -- ISM: magnetic fields -- polarization -- radio lines: ISM}

\section{Introduction}\label{sect-intro}

\subsection{Gas Inflow in the Galactic Center}

The origin of the 2-pc circumnuclear disk (CND) in the Galactic Center (GC) has remained unclear in spite of intensive studies over the past decades \citep[e.g.,][]{guesten87,jackson93,amo11,harris,mezger,etx,lau,dent93,morris96,wright,maria,martin12,herrnstein02, herrnstein05,chris,great,mills13b,hsieh17}.
The CND is a ring-like molecular structure rotating with respect to the supermassive black hole (SMBH) SgrA*, within which are the  ionized gas streamers called SgrA West (mini-spiral) \citep{roberts93,irons12,lacy91,wang10,paumard04,sco03,zhao09,lo83,tsuboi17,tsuboi16}.
The CND, being the closest molecular reservoir in the GC, is critical on the understanding of the feeding of the nucleus. The replenishment of the CND itself, therefore, is an important problem.

Previous studies have shown that multiple dense gas streamers surrounding the CND may carry gas directly toward the nuclear region. Parts of these streamers might be captured by the central potential \citep{okumura89,ho91,coil00,mcgary01,hsieh17}. In particular, our analysis of the gas kinematics in the CND and streamers presented in \citet{hsieh17} indicates that these streamers show a signature of rotation with an additional inward radial motion with progressively higher velocities as the gas approaches the CND and finally ends up co-rotating with the CND. Our results  suggest a possible mechanism of gas feeding towards the CND from 20 pc to around 2 pc. We also find that the western and  eastern streamers can be described by a simple model of Keplerian rotation and infall. The western streamer also shows kinematic and morphological evidence of an intersection with the CND, possibly supplying  gas to the CND.
Within the CND, the parts with densities lower than the tidal threshold will be drawn inwards to the GC.

The magnetic field ($B$-field) is expected to be important for material orbiting within the CND \citep{aitken86,werner88,hilde90,hilde93} as the $B$-field can not only make the gas resist more against gravity but also helps accretion by removing angular momentum. Moreover, investigating the $B$-field in the GC is necessary because the energy stored in the $B$-field is expected to be comparable to the gas kinetic, the radiation, and the cosmic ray energies, if the equipartition theorem is valid \citep{cf53}. The observed mG $B$-field \citep{plante94} at a $100$-pc scale is seen  to interact with the molecular gas, as indicated by the molecular loop \citep{fukui06}, the nonthermal radio filaments (NTFs) \citep[e.g.][]{yusef84,yusef04,larosa04}, and the Galactic Center arc \citep{serabyn87,serabyn94,tsuboi97}.

\begin{figure*}[bht]
\begin{center}
\includegraphics[angle=0,scale=0.4]{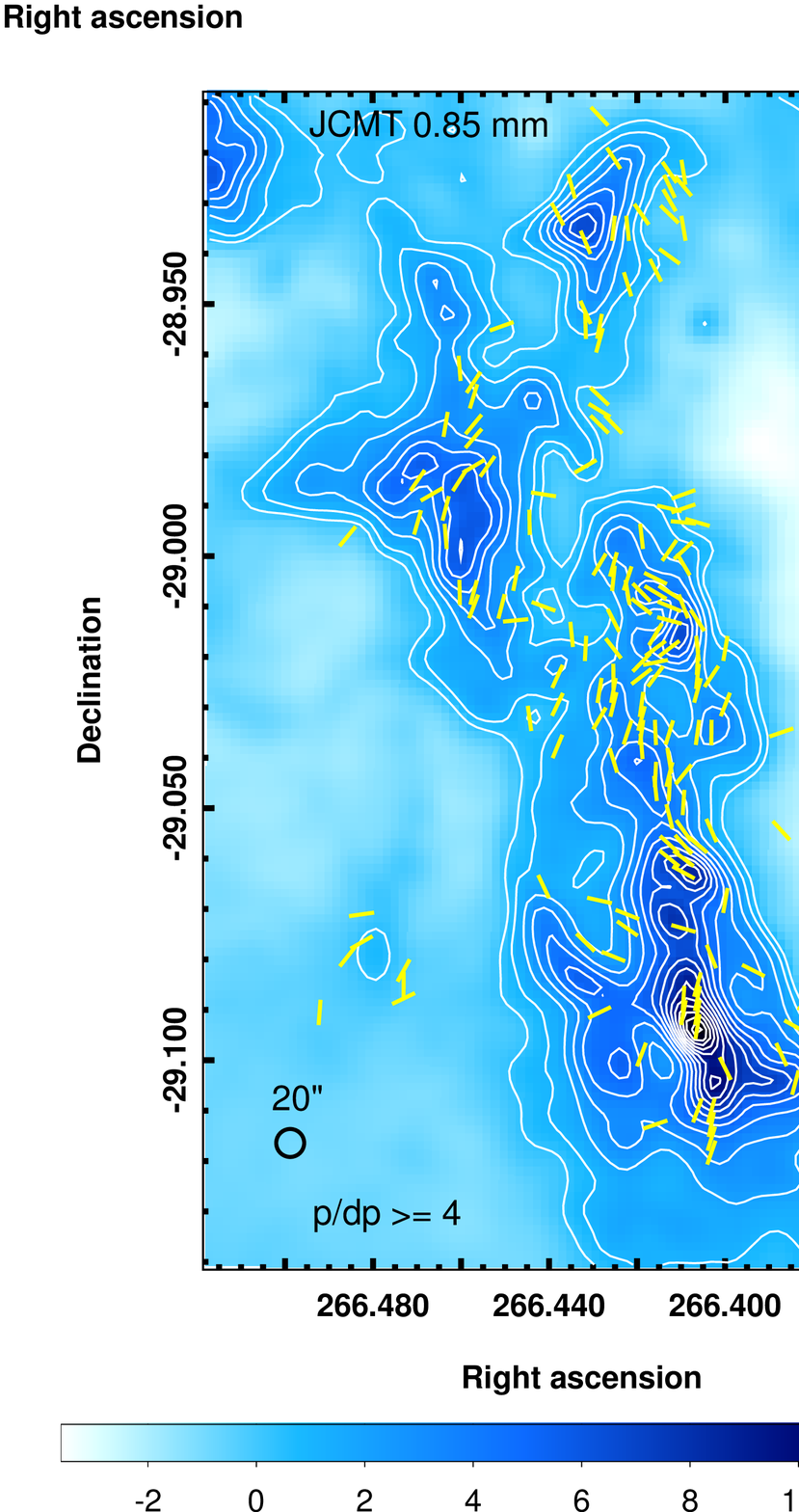}
\caption[]{\linespread{1}\selectfont{} Color map shows the SCUBA 850 $\micron$ continuum from \citet{francesco08}.
Color map shows the SCUBA 850 $\micron$ continuum from \citet{francesco08}, with solid contours spaced at 0.35, 1, 2, 3, $\cdots$, 14 Jy beam$^{-1}$ with an interval of 1 Jy beam$^{-1}$.
The beam size is 20$\arcsec$. The SCUPOL polarization segments (yellow) show the orientation of the  magnetic field in $p/dp~(=\sigma_{\rm P}) \ge2, 3, 4$ in the top left, top right, and bottom panel, respectively.  The 450 $\micron$ $B$-field  detected with SPARO \citep{dotson98,novak03} with a resolution of 6$\arcmin$ is overlaid with red segments in the top leftpanel. The regions encompassing the 50 MC, the 20 MC, and the CND --  magenta boxes in the top right panel -- are compared with the 350 $\micron$ Hertz data in Appendix A (Figure~\ref{fig-compare}).}
\label{fig-frans}
\end{center}
\end{figure*}

\begin{figure*}[ht]
\begin{center}
\epsscale{0.6}
\includegraphics[angle=0,scale=0.8]{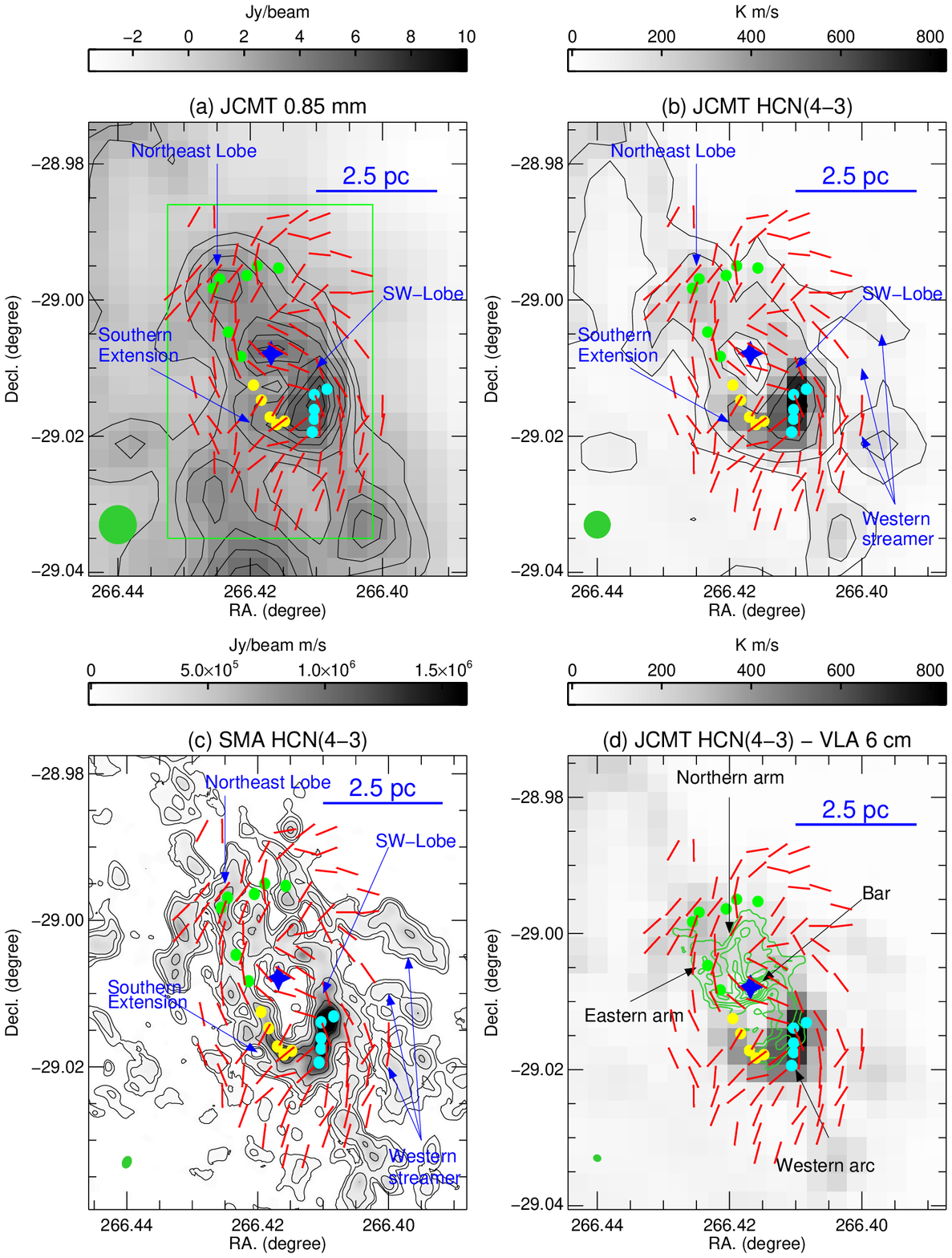}\vspace{-1.75cm}
\caption[]{\linespread{1}\selectfont{}\small
(a): SCUPOL $B$-field (red segments) overlaid on the 850 $\micron$ continuum map. The beam size of 20$\arcsec$ is  shown with a green circle.  The segments are sampled on a 10$\arcsec$ grid. Contour levels are 20$\%$, 30$\%$, ..., 100$\%$ of the intensity peak (6.5 Jy beam$^{-1}$). We show the $B$-field data associated with the CND (the green box: $1.86\arcmin\times2.9\arcmin$). The $B$-field angle is measured north of west.
(b): $B$-field segments overlaid on the JCMT HCN($J=4-3$) map (beam: 14$\arcsec$).
Contour levels are 10$\%$, 20$\%$, 40$\%$,.., 100$\%$ of the peak (835 K m s$^{-1}$).
(c): $B$-field segments overlaid on the SMA HCN($J=4-3$) map \citep{maria} (beam:$5.9\arcsec\times4.4\arcsec$).
Contour levels are 57, 115, 229, 459, 918, 1376, 1835, 2294 Jy beam $^{-1}$ km s$^{-1}$.
Earlier identified features in the CND -- northeast lobe (green dots), southwest lobe (cyan dots), and southern extension (yellow dots), and western streamer are labeled.
(d): $B$-field segments overlaid on the JCMT HCN($J=4-3$) map (color) and the VLA 6 cm continuum (contours; beam: $3.8\arcsec\times3.2\arcsec$).
Contour levels are 0.07, 0.1, 0.157, 0.246, 0.335, 0.424, 0.51, 0.6, 0.69, 0.78, 0.87, 0.95, 1 Jy beam$^{-1}$ (beam: $3.8\arcsec\times3.2\arcsec$ ).
SgrA* is labeled with the blue star. The eastern arm, northern arm, bar, and western arc of the mini-spiral are labeled.}
\label{fig-vgb}
\end{center}
\end{figure*}

The large-scale (100 pc) configuration of the $B$-field in the GC region consists of two structures identified from NIR \citep{nishiyama10} and submillimeter polarization data \citep{novak00}.
On a large scale, the $B$-field shows a vertical structure (poloidal) in the NIR polarimetry above 0.4$\degr$ off the Galactic plane.
Near the Galactic plane,  the field orientation gradually becomes toroidal \citep{dotson98,novak03}.
In the finer structures revealed by the 350 $\micron$ Hertz data at a $20\arcsec$ resolution, the B-field lines tend to follow the long axis of the molecular clouds \citep{chuss03}.
The 350 $\micron$ flux shows a linear and positive correlation with the $B$-field position angles \citep{chuss03}, which suggests that the field is more poloidal in less dense regions while the field is more toroidal in the denser regions. This correlation was attributed to the tidal shear in the central molecular zone (CMZ) \citep{aitken91,aitken98,chuss03}.
Moreover, near-infrared polarimetric data have shown that the orientation of the $B$-field is more  parallel to the Galactic plane (toroidal), but is nearly  perpendicular to the plane above $b\ge0.4^{\circ}$ \citep{nishiyama10}. This provides evidence for a smooth transition of the large-scale $B$-field.
\citet{uchida85} constructed a model to connect poloidal and toroidal field components based on the assumption of magnetic flux-freezing.  They found that differential rotation can shear an initially poloidal field into a toroidal field in sufficiently dense regions.

Compared to the large-scale Galactic disk, the configuration and the role of the $B$-field in the CND are not well understood despite the B-field's potential dynamical importance in e.g., the gas inflow that can feed the SgrA* and its surrounding, the CND \citep{hilde90,hilde93,greaves02}, the mini-spiral \citep{aitken91,zylka95,aitken98,aitken00}, and the collimation of outflows \citep[e.g.,][]{zhao14,hsieh16}. Furthermore, recent simulations \citep{blank16} have also shown that  a mG-$B$-field can support the stability of  the  ionized cavity for at least 10$^{6}$ yr.

To date, there are only a few attempts to construct models of the $B$-field configuration in the CND. In particular, \citet{wardle90,desch97} investigated a smooth disk model for the CND including the $B$-field.  They adopted an axially symmetric magnetic model where an initially poloidal field threading the differentially rotating disk is pulled into a toroidal
configuration. Their models can describe the 100 $\micron$ polarization data reasonably well \citep{hilde90,hilde93}.
In their model, the azimuthal ($B_{\phi}$) and the radial ($B_{\rm r}$) components exceed the vertical ($B_{\rm z}$) component which implies that the field can remove the excess angular momentum of the accreted material \citep{wardle90}.

The CND is undoubtedly  a key object for investigating the interplay between gas dynamics and the magnetic field near SgrA*.
Yet, the effects of the $B$-field on the gas inflow associated with the CND, its streamers, and the inner mini-spiral are not clearly understood at parsec-scale. Previous polarization measurements toward the CND have mostly been reported  at shorter wavelengths (350 $\micron$ and 100 $\micron$), while the polarization from cooler components at 850 $\micron$ has not yet been probed in detail \citep[see][]{matthews09,aitken00}.
In this paper, we investigate the $B$-field configurations and their relationship  with the molecular gas in the GC using the 850 $\micron$ SCUPOL archival data from the James Clark Maxwell Telescope (JCMT) \citep{matthews09}.
We will inspect the polarized emission at 850 $\micron$ at a sub-parcsec resolution (20$\arcsec=0.76$ pc) toward the CND.


\begin{figure*}
\begin{center}
\epsscale{1}
\vspace{-1.5cm}
\includegraphics[angle=-90,scale=0.76]{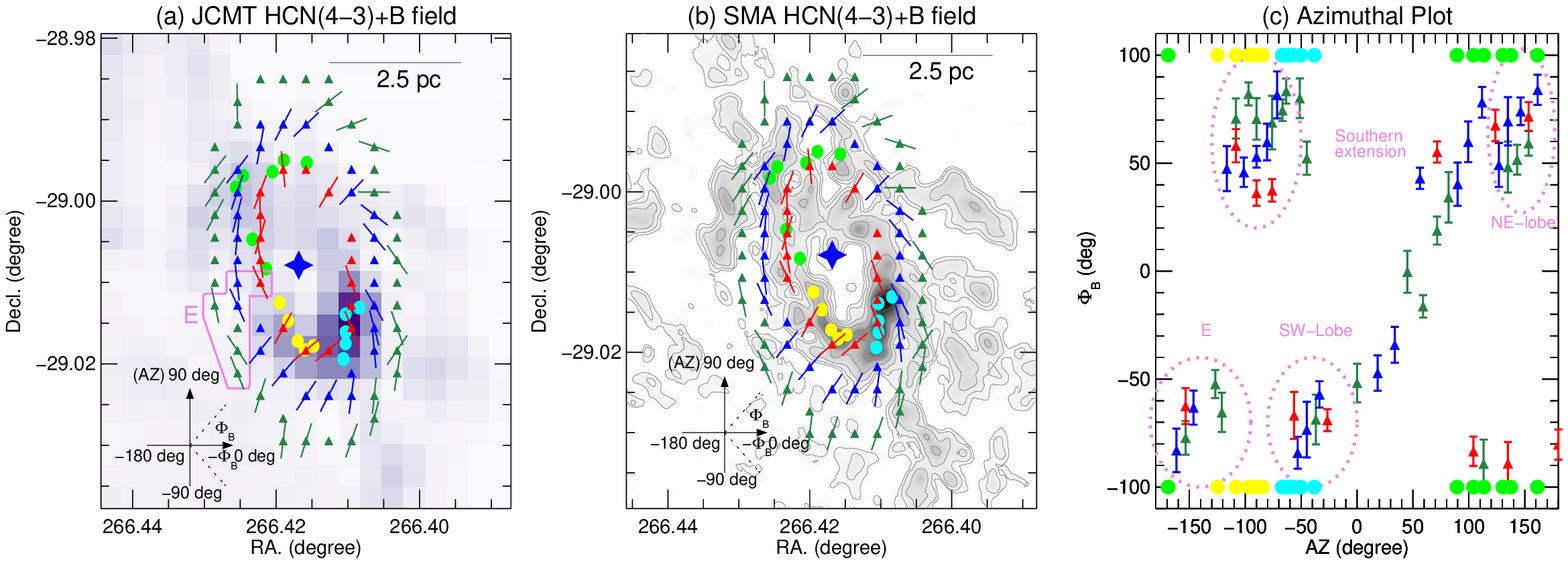}\vspace{-1.25cm}
\caption[]{
(a): Azimuthal variations of the 850 $\micron$ $B$-field lines (green, blue, and red segments).
Sampled elliptical annuli are indicated with green, blue, and red triangles labelling the outer, middle, and inner annulus, respectively. The center of the ellipses is SgrA* (blue star). The background color image is the archival JCMT HCN($J=4-3$) map. 
(b): Same as (a), but the gray/contour map shows  SMA HCN($J=4-3$). The contour levels are the same as Figure~\ref{fig-vgb}c.
(c): $x$- and $y$-axis are the azimuthal angle ($az$) and the position angle of the $B$-field segments ($\Phi_{\rm B}$), respectively. The reference ($0\degr$) of $az$ and $\Phi_{\rm B}$ is west. $\Phi_{\rm B}$ is shown from $-90\degr$ to $90\degr$ (see the definition for the range of coordinate in the lower left corner in (a) and (b)). The green, blue, and red triangles mark the $\Phi_{\rm B}$ sampled along the green, blue, and red annulus shown in (a) and (b). The locations of the NE-lobe (green dots) , SW-lobe (cyan dots), and southern extension (yellow dots) are labeled in (b) and (c). The region ``E'' and its corresponding $B$-field segments are marked in (a), revealing a discontinuity in orientations from E to the southern extension. 
}
\label{fig-bf-vg}
\end{center}
\end{figure*}


\subsection{ 850 $\micron$ SCUPOL Polarization Archival Data}

In Figure~\ref{fig-frans}, we show the archival 850 $\micron$ polarization  data from the legacy program  of SCUPOL with the JCMT \citep{matthews09} (effective 20$\arcsec$ beam$\sim0.76$ pc). \citet{matthews09} collected the JCMT data toward the GC, where the $B$-field is sampled on a 10$\arcsec$ grid.
The linearly polarized light from dust grains is frequently used to probe the integrated plane-of-sky $B$-field morphology. Interstellar dust grains are elongated, with their minor axes parallel to the $B$-field. The thermal emission from the aligned dust grains is then polarized with polarization segments perpendicular to the field lines \citep{cudlip82,hilde84,hilde88,laza00,ander15}. The dust polarization can, therefore, reveal the plane-of-sky projected $B$-field orientations.
In Figure~\ref{fig-frans}, the SCUPOL $B$-field segments are overlaid on the JCMT 850 $\micron$ map \citep{francesco08}. The segments are plotted with $p/dp~(=\sigma_{\rm P}) \ge2, 3, 4$. The 450 $\micron$ Submillimeter Polarimeter for Antarctic Remote Observing (SPARO) $B$-field \citep{novak03} is also overlaid with a resolution of 6$\arcmin$ (corresponding to a linear scale of 13.7 pc). The low-resolution SPARO map traces the large-scale $B$-field  which is parallel to the plane of our Galaxy. This alignment of  field orientations with the Galactic plane is attributed to a large-scale toroidal $B$-field configuration (i.e., azimuthal field). 
The SCUPOL 850 $\micron$ $B$-field as well as the continuum are clearly detected and resolved along and across the sub-features in the GC at a 20$\arcsec$ resolution ($=0.76$ pc).  The detected $B$-field orientations ($\Phi_{\rm B}$) vary enormously over the entire map, ranging from $-90\degr$ to $90\degr$ (0$\degr$ is west, positive is counterclockwise). Nevertheless, the $\Phi_{\rm B}$ varies smoothly and systematically along certain sub-structures, revealing organized patches in, e.g., the CND, the giant molecular cloud 20/50 MC \citep[e.g.][]{gusten81}, and the HVCC CO 0.02-0.02 \citep{oka99}.
The azimuthal correlation seen in the CND is providing the link between the model proposed by \citet{wardle90} and the SCUPOL 850 $\micron$ polarization data.  We are exploring this link in the following sections. 
A comparison of polarization data is presented in Appendix A.
We find that the morphology of the $B$-field is consistent with different threshold cuts in $\sigma$. 
Hence, in order to work with the maximum of independent data points, we present data with $\sigma_{\rm P} \ge2$, and we will focus on the $B$-field structure of the CND.

\section{$B$-field of the CND traced with 850 $\micron$ Polarization}
\subsection{ Structures of CND and $B$-field}

Figure~\ref{fig-vgb}  displays the JCMT 850 $\micron$ continuum map of the CND (green box: central $1.86\arcmin\times2.9\arcmin$) with the $B$-field lines overlaid. The CND consisting of the northeast lobe, SW-lobe, and southern extension is surrounded by the 20/50 MC. Compared to the 850 $\micron$ continuum, the CND shows a more prominent ring-like feature in the high-excitation HCN($J=4-3$) line surrounded by several streamers \citep{hsieh17}.
The JCMT HCN($J=4-3$) data are from the JCMT archive (project code: M06BC06). We reduced the data using the ORAC-DR pipeline software. The $B$-field lines show complicated but organized structures associated with the CND. In order to reveal more smaller-scale features that might help to explain the $B$-field morphology, we additionally present a higher-resolution SMA HCN($J=4-3$) map (Figure~\ref{fig-vgb}c).
Known features of the CND and its surrounding \citep{maria,martin12} are also labeled in Figure~\ref{fig-vgb}: (1) northeast lobe (NE-lobe), (2) southern extension, and (3) southwest lobe (SW-lobe). These three main features are fitted  as three Keplerian components with the CN($J=2-1$) line \citep{martin12}. We also label these 3  kinematic components in Figure~\ref{fig-vgb} and \ref{fig-bf-vg} with green, yellow, and cyan dots. The positions of these dots are determined by \citet{martin12}.

A detailed look at the position angles of the $B$-field, $\Phi_{\rm B}$, reveals systematic changes in $\Phi_{\rm B}$ along individual streamers or filaments at pc-scale. Overall, the $\Phi_{\rm B}$ associated with the CND can be described by at least three azimuthal components, i.e., the NE-lobe, the southern extension, and the SW-lobe (Figure~\ref{fig-vgb}b). The $B$-field  shows smooth radial variations along the long axes of individual streamers. The locations of these three $B$-field components spatially coincide with the three Keplerian components mentioned above. Besides, several streamers surrounding the CND were reported to feed the CND \citep{okumura89,ho91,coil00,mcgary01,hsieh17}. These streamers are the SW-Lobe and western streamer, which  show signatures of rotation and inward radial motion with progressively higher velocities as the gas approaches the CND and finally ends up co-rotating with the CND \citep{hsieh17}.

In Figure~\ref{fig-vgb}d, we also show the archival  Very Large Array (VLA) 6 cm continuum tracing the mini-spiral (SgrA West) \citep[e.g.,][]{roberts93}.
The mini-spiral consists of the arc-shaped ionized gas streamers converging onto  SgrA*. These streamers have been proposed to originate from the inner edge of the CND.  The western arc of the mini-spiral might be the boundary of the CND, ionized by the central cluster \citep{martin08,amo}. This association is kinematically consistent with the HCN($J=1-0$) and H92$\alpha$ lines \citep{chris}.
Very interestingly, Figure~\ref{fig-vgb}d reveals 
that around the mini-spiral, the $B$-field lines curve along the arcs of the mini-spiral. This may hint an underlying coherent magnetic structure that is connecting the CND and the mini-spiral. We will discuss their link in the later sections.

\subsection{Azimuthal Correlation of $\Phi_{\rm B}$}

To further elucidate the $B$-field structures in the CND, we are looking at the azimuthal correlation of $\Phi_{\rm B}$ in the CND in Figure~\ref{fig-bf-vg}.
The azimuthal angles ($az$) and the $B$-field position angles ($\Phi_{\rm B}$) are both measured counterclockwise, starting from west.
The $az$ angles are measured for the $B$-field segments selected along the three concentric ellipses, marked by the green, blue, and red triangles. The center of the ellipses is SgrA*, where the ellipses are oriented with their major axes aligned with the major axis of the CND.
We find that $\Phi_{\rm B}$ varies systematically as a function of $az$ along the elliptical annuli (Figure~\ref{fig-bf-vg}c). The $\Phi_{\rm B}$ of the southern extension of the CND (yellow dots; $az\sim-100\degr$) is $\sim70\degr$.  The $\Phi_{\rm B}$ then shows a smooth variation from the SW-lobe (cyan dots; $az\sim-60\degr$) to the NE-lobe (green dots; $az\sim100\degr$). This figure demonstrates a clear azimuthal structure in $\Phi_{\rm B}$ which connects the various features along the CND.
We note that there is an apparent sudden jump in $\Phi_{\rm B}$ from the region E to the southern extension. 
We have to acknowledge that the representation with $\Phi_{\rm B}$ can be ambiguous for values close to $\pm 90\degr$ which can mimic a discontinuity in $\Phi_{\rm B}$. Nevertheless, the sudden jump in $\Phi_{\rm B}$ between the region E and the southern extension appears real as can be seen in particular for the inner red and blue segments where a change in orientations of about $90\degr$ is detected. 
The close-to-axisymmetric pattern in $\Phi_{\rm B}$ in Figure~\ref{fig-bf-vg} resembles the self-similar $B$-field model presented by \citet{wardle90} (hereafter WK-model), as we will analyze further in Section 3.

Finally, we note that the lower-resolution 100 $\micron$ data (beam$=45\arcsec$) in \citet{hilde90,hilde93} already
revealed a trend. The first observations in \citet{hilde90} contained only six data points with polarization position 
angles mostly along an east-west direction in a "dust ring" around Sgr A*. The later observations in \citet{hilde93} tripled the number of data points with new detections in the western region that started to hint a rotation 
in position angles, providing a clear indication for an axisymmetrical model. The high-resolution 850 $\micron$ (beam$=20\arcsec$) SCUPOL data presented and analyzed here 
reveal a dramatically improved polarization coverage with significantly more structural details
as compared to the earlier works in \citet{hilde90,hilde93}. This is the reason why we are focusing on the
850 $\micron$ data in our comparison with the WK-model in the following section.
Appendix A gives further details on the comparison among different polarization data sets. Similar to the 100 $\micron$ map, the 350 $\micron$ data show a $B$-field with a north-south orientation. In any case, substantially more detailed sub-structures are resolved with the 850 $\micron$ polarization data analyzed here.


\begin{figure*}[ht]
\begin{center}
\epsscale{0.4}
\includegraphics[angle=0,scale=0.7]{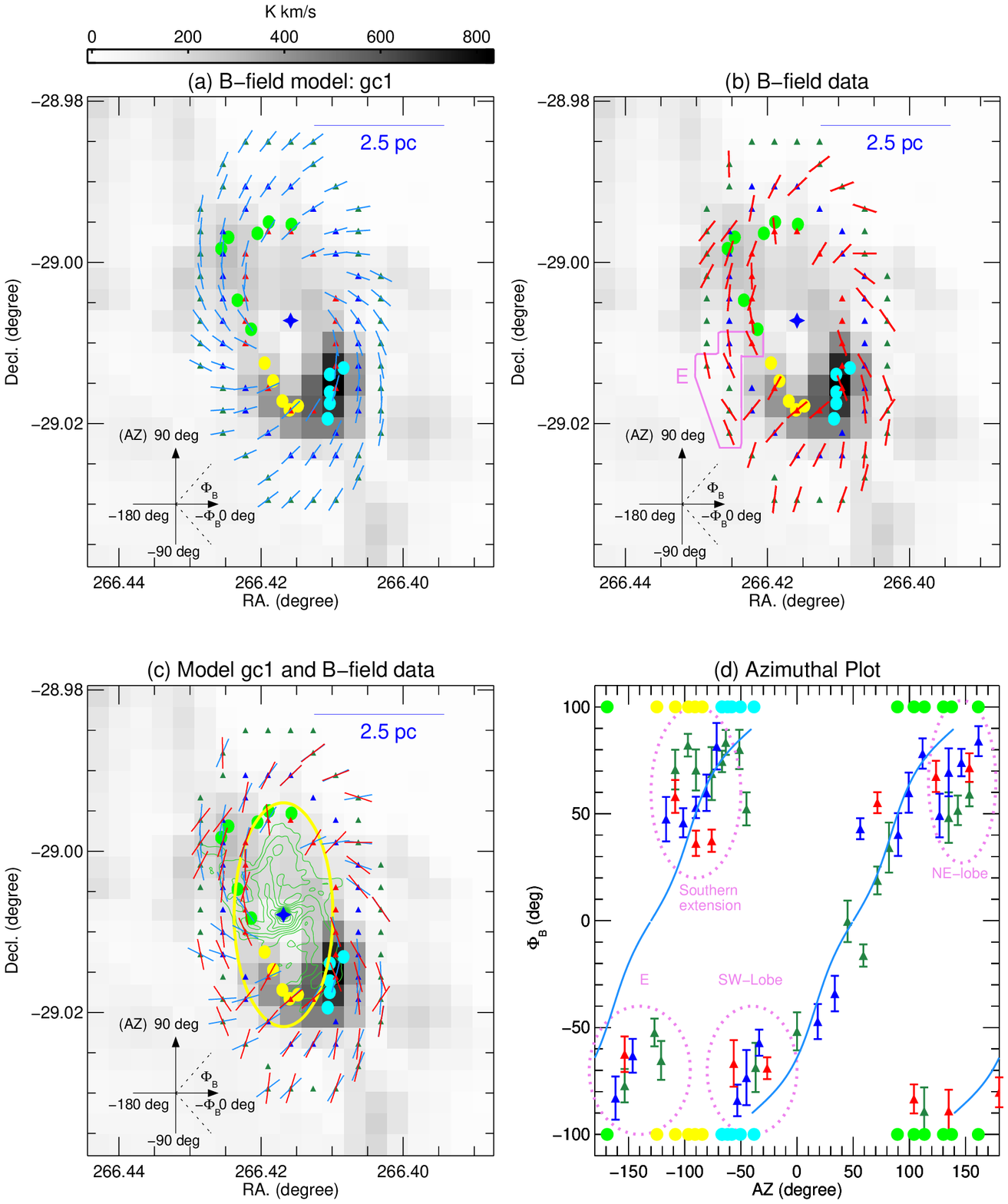}\vspace{-1cm}
\caption[]{\linespread{1}\selectfont{}
(a): the $B$-field model (gc1) in the CND is shown in blue segments overlaid on the JCMT HCN($J=4-3$) map. The $B$-field is sampled along the same annuli (red, blue, and green triangles) as shown in Figure~\ref{fig-bf-vg}. A scale bar of 2.5 pc is shown with the blue line.
(b) The SCUPOL $B$-field data are shown with red segments.
Note that 74\% of the segments are $\sigma_{\rm P}\ge3$ in the CND and the mean uncertainty is $\sim7.5^{\circ}$.
(c) The $B$-field model (blue segments) and the data (red segments) are overlaid on the HCN($J=4-3$) map (color) and the VLA 6 cm continuum contours (the contour levels are the same as Figure~\ref{fig-vgb}d.). The yellow ellipse marks the boundary of the mini-spiral.
(d) $az$-$\Phi_{\rm B}$ correlation of the three annuli (red, blue, and green triangles). Error bars of $\pm1\sigma$ are shown. The $B$-field model (gc1) is overlaid with the blue line.
}
\label{fig-model-az-gc1}
\end{center}
\end{figure*}

\section{Comparison with the WK-Model}

\subsection{Possible $B$-Field Configurations}

In the WK-model, the open $B$-field lines originally perpendicular to the Galactic plane are  
wrapped by the rotating radial inflow, and thus the $B$-field lines within the Galactic plane are parallel to the plane. The $B$-field structure is axially symmetric and the predicted orientation of polarization is the same at a constant radius. 
This self-similar magnetized disk model with an axially symmetric configuration gives a satisfactory fit to the far-infrared 100 $\micron$ polarization of the CND \citep{hilde90,hilde93}.
Here, we adopt the 5 self-similar disk models used in \citet{wardle90}, called gc1, gc2, gc3, gc4, and gc5. These 5 models are representative cases of varying $B_{\rm r}$ (radial $B$-field component), $B_{\rm z}$ (vertical $B$-field component), and $B_{\rm \phi}$ (azimuthal $B$-field component). Four dimensionless  parameters specify each of these 5 models. They are $\alpha$, $\beta$, $\delta$, and $\epsilon$ \citep[also see Figure 2 and Table 1 in][]{wardle90}. $\alpha$ and $\beta$ fix the ratios of $\left|B_{\phi}/B_{\rm z}\right|$ and $\left|B_{\rm r}/B_{\rm z}\right|$, respectively. $\delta$ describes the thickness of the disk in a thin-disk approximation ($\delta\ll1$). $\epsilon$ is the ratio between inflow and azimuthal velocity ($v_{\rm r}$ and $v_{\phi}$).
The main features of these models are as following. A bar indicates averages along $z$ through the disk. We note that the ratio $\epsilon$, the inflow-to-rotational velocity,  is 0.1 for gc2, and 0.25 for all the other models \citep{wardle90}.
\begin{itemize}

\item{gc1: represents the case where $\overline{\left|B_{\rm r}\right|}\approx\overline{\left|B_{\phi}\right|}>\overline{B_{\rm z}}$ ($\alpha=40$, $\beta=27.0$).}
\item{gc2: $\epsilon$ is reduced from 0.25 to 0.1. $\overline{\left|B_{\rm r}\right|}<\overline{\left|B_{\phi}\right|}$, but $\overline{\left|B_{\rm r}\right|}>\overline{B_{\rm z}}$ is held ($\alpha=40$, $\beta=16.9$).}
\item{gc3: $\overline{\left|B_{\rm z}\right|}\gtrsim\overline{\left|B_{\phi}\right|}$ and $\overline{\left|B_{\rm r}\right|}$ is larger than $B_{\rm z}$ ($\alpha=15$, $\beta=16.5$).}
\item{gc4: $\left|B_{\phi}\right|$ remains unchanged and $\left|B_{\rm r}\right|$ is smaller than $B_{\rm z}$ everywhere ($\alpha=40$, $\beta=7.5$).}
\item{gc5: $\overline{B_{\rm z}}$ remains larger than $\overline{\left|B_{\phi}\right|}$ and $\overline{\left|B_{\rm r}\right|}$ everywhere ($\alpha=15$, $\beta=7.5$).}

\end{itemize}

The predicted polarization angle $\Phi_{\rm p}$ ($\chi$ in the original paper) of the integrated $B$-field is given by the following equations \citep{wardle90}:

\begin{equation}
\cos{2\Phi_{\rm p}}=\frac{q}{(q^2+u^2)^{1/2}},~\sin{2\Phi_{\rm p}}=\frac{u}{(q^2+u^2)^{1/2}}.
\end{equation}

The Stokes parameters $q$ and $u$ are

\begin{equation}
q/N_{\rm d}=\sin^{2}{\omega}[\cos^{2}{\xi}(\cos^{2}{\xi}(\cos^{2}{i}+1)-1)]+\cos^{2}{\omega}\sin^{2}{i},
\end{equation}
\begin{equation}
u/N_{\rm d}=\cos{i}\sin^{2}{\omega}\sin{2\xi},
\end{equation}
\begin{equation}
\xi=\theta+az,
\end{equation}
where $\theta$ is the angle between the outward cylindrical radius vector and the projection of $B$ in the upper layer onto the mid-plane \citep[see Figure 4b in][]{wardle90}. $N_{\rm d}$ is the column density of dust, $\omega$ is the angle between $B$ and the $z$-axis, and 
$az$ is the azimuthal angle measured from the west in the plane of the sky. $i = 70^\circ$ is the
inclination of the disk, adopted from \citet{guesten87}.
The resulting projected polarization angle $\Phi_{\rm p}$ can then be shown as a function of $az$. 
The projected $B$-field lines for these 5 self-similar models -- rotated by $90\degr$ with respect to the derived polarization orientations -- are shown in Appendix B (Figure~\ref{fig-4model}). We remark that in \citet{wardle90}, \citet{hilde90}, and \citet{hilde93}, $\Phi_{\rm p}$ is measured east of north, while in our paper, the $\Phi_{\rm p}$ is measured north of west. Therefore, the $\Phi_{\rm p}$ determined in the above equations directly corresponds to the angle of the $B$-field, assuming that the polarization angle ($E$-field) is perpendicular to the $B$-field for dust polarization observed in the submillimeter at our densities and temperatures. 
Hence, in the following we will compare the observed position angles of the $B$-field, $\Phi_{\rm B}$, with the ones from the model, $\Phi_{\rm model}\equiv\Phi_{\rm p}$.

In Appendix B, we present the projected $B$-field configurations predicted by the WK-models in comparison with the SCUPOL 850 $\micron$ observations. We further analyze the differences, $\Phi_{\rm B}$-$\Phi_{\rm model}$, between the predicted and observed angles as a function of $az$. The differences are minimized for model gc1, suggesting that the magnetic field in the CND is equally dominated by $B_{\rm r}$ and $B_{\phi}$ with an inflow-to-rotational velocity ratio
of 0.25. In what follows, we adopt model gc1 as the best-fit model and compare it further with the observations. 
The reader is referred to Appendix B for a detailed comparison of the other models.

The result of the predicted $\Phi_{\rm model}$  of gc1 as a function of $az$ is shown in Figure~\ref{fig-model-az-gc1} (other models are shown from Figure~\ref{fig-model-az-gc2} to \ref{fig-model-az-gc5} in Appendix B). The model (panel a) and data points (panel b)  are sampled within the same elliptical annuli as shown in Figure~\ref{fig-bf-vg}. 
Panels c and d indicate that the $B$-field data can in general be described by the model gc1 (blue line in panel d) within $\pm1\sigma$ uncertainties.
However, in the eastern region (marked as E in panel d), deviations between data and model ranging from $20\degr-60\degr$ are found. The largest deviations are seen in the outer ring (green triangles). 
The SW-Lobe has deviations from 10$\degr$ to 20$\degr$, and the NE-lobe (NE-lobe in panel d) shows deviations from 20$\degr$ to 40$\degr$. In fact, none of the models can give a single fit to all the data points. The model gc3 is able to fit the southwest-northeast transition along the CND, but some data points (around $az=0\degr-60\degr$ and in region E) show large deviations around $\ge30\degr$  Lastly, the models gc4 (dominating $B_{\phi}$) and gc5 (dominating $B_{\rm z}$) are less accurate descriptions for their larger and systematic offsets from the data.

\subsection{Spatial Correlation of CND, Mini-Spiral, and $B$-field}

\begin{figure*}[ht]
\begin{center}
\epsscale{0.5}
\includegraphics[angle=-90,scale=0.7]{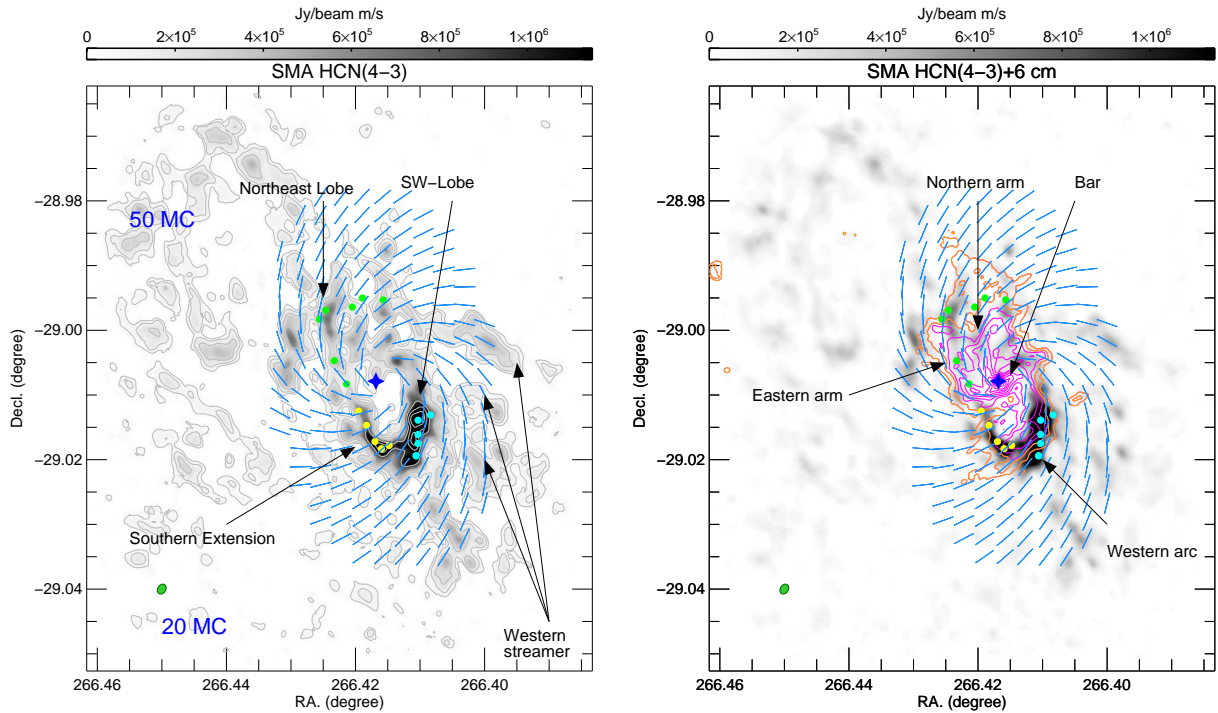}\vspace{-0.5cm}
\caption[]{
Left: $B$-field model of gc1 (blue segments) overlaid on the SMA HCN($J=4-3$) map.
Right: $B$-field model of gc1 (blue segments) overlaid on the SMA HCN($J=4-3$) map (gray) and the VLA 6 cm contours (orange and magenta). The known features of the mini-spiral (western arc, bar, northern arm, and eastern arm) are labeled. The contour levels are 0.03, 0.05, 0.07, 0.1, 0.157, 0.246, 0.335, 0.424, 0.51, 0.6, 0.69, 0.78, 0.87, 0.95, and 1 Jy beam$^{-1}$. The model $B$-field orientations are aligned with the streamers and the CND, which suggests that the same field lines (gc1), continuing inward, also align with the three arms of the mini-spiral.
}
\label{fig-model-6cm}
\end{center}
\end{figure*}

In Figure~\ref{fig-vgb}d we have found that the SCUPOL 850 $\micron$ $B$-field lines seem to connect to and morphologically match the mini-spiral with its northern arm, its eastern arm, and its western arc. To further investigate the possible association between the CND, the mini-spiral and the $B$-field,
we overlay the best-fit model gc1 -- identical to the model used in Section 3.1 where the most inner part was not displayed -- on the SMA HCN($J=4-3$) and the 6 cm maps in Figure~\ref{fig-model-6cm}.  As argued in the previous section, the model $B$-field orientations are aligned with the streamers and the CND. Figure~\ref{fig-model-6cm} suggests that the same field lines, continuing inward, also align with the three arms of the mini-spiral.
In Figure~\ref{fig-model-data-6cm} we display the $az-\Phi_{\rm B}$ plot of the data and model (gc1) in the inner region of the CND. Similar to the CND, gc1 also gives the best Chi-square fit ($\chi^{2}$) in the inner region ($\chi^{2}/dof=17.48$, degree of freedom $(dof)=$21,  $P=68\%$). This result provides some evidence that the CND and the mini-spiral are  connected through the same coherent $B$-field,  such that the morphology of the mini-spiral follows the pattern of the $B$-field. Nonetheless, in the western arc, there are two points (SW1, SW2) that have large deviations ($\ge80\degr$) from the model (panel d in Figure~\ref{fig-model-data-6cm}) . We will discuss these deviations further in Section 4.


\section{Discussion}

In the previous section we have found that the $B$-field in the CND has comparatively important $B_{\phi}$ and $B_{\rm r}$ components (gc1), while $B_{\rm z}$ is less important. We have further suggested that the best-fit model gc1 can also describe the $\Phi_{\rm B}$ of the mini-spiral. A detailed comparison between the other models and filaments of the CND is presented in the following. 

\subsection{The CND -- A Magnetized Disk}

As already hinted by the earlier lower-resolution 100 $\micron$ polarization data \citep{hilde90,hilde93},  the $B$-field of the CND traced by the higher-resolution 850 $\micron$ SCUPOL data at a $20\arcsec$ resolution can be described as an axially symmetric field, which is generated by the differentially-rotating radial inflow in the WK-model. The initial Galactic poloidal field is stretched and wrapped  by the inflows with rotational shear to produce the observed azimuthal field.
This scenario is also proposed based on the measurements done by \citet{chuss03} who find that the initial poloidal field is collapsed to the denser regions near the Galactic plane.
The $B$-field of the CND seems to be predominately azimuthal ($B_{\rm \phi}$).
However, the radial component ($B_{\rm r}$) cannot be ignored when compared
to $B_{\rm \phi}$. The extreme cases of gc4 ($B_{\phi}$ largely dominates) and gc5 ($B_{\rm z}$ largely dominates) are less likely the dominant case in the CND. In the previous sections, we have found that gc1 gives the best fit to the 850 $\micron$ polarization data. A 25$\%$ deviation from circular rotation is required in gc1 
(inflow-to rotational velocity ratio $\epsilon=0.25$). 
This is similar to the value we derived in the previous studies where we constrained the radial inward motion to be about 20-40$\%$ of the circular rotation in the CND and surrounding streamers, based on a phase plot analysis \citep{hsieh17}. Besides, the neutral streamers are likely converging to the CND. These streamers are likely to follow the field lines as predicted in the WK-model shown in Figure~\ref{fig-model-6cm}. 
However, the detailed comparison in Appendix B also indicates that model gc2 cannot be rejected.
The difference between gc1 and gc2 is that the inflow velocity in gc2 (10$\%$ deviation from circular rotation) is smaller than in gc1. This leads to a reduced inward magnetic flux advection, resulting in $B_{\rm r}$ being smaller than $B_{\phi}$. 

We note that \citet{hilde90} suggested gc1 to be the best-fit configuration for only six mostly north-south oriented 
$B$-field segments from their lower-resolution 100 $\micron$ polarization data.  The later data set with about three
times more data points in \citet{hilde93} still revealed a prevailing north-south orientation of the $B$-field but with some additional inward bending. This later work favored the axisymmetrical model gc2.
The $B$-field morphology traced at 850 $\micron$ shows dramatically more sub-structures and a much larger 
coverage than at 100 $\micron$. 
The observations at shorter wavelengths (100 $\micron$, 350 $\micron$, and 450 $\micron$) possibly trace warmer dust components, while the 850 $\micron$ emission is dominated by a cooler and likely denser dust component, i.e., the CND and inflow material. Hence, polarization emission coming from different regions may lead to different interpretations when compared to the models. Higher-resolution observations, such as e.g., with SOFIA at near-infrared wavelengths,  will provide further important clues on the separation and origin of different dust components.

We are returning to the SW-lobe in Figure~\ref{fig-model-az-gc1}d
which displays $\sim20\degr$ deviations from gc1.
The complication in the SW-lobe is likely due to the presence of incoming material that is accreted onto the CND  \citep{hsieh17}.
We find that gc3 can fit the SW-lobe better than gc1. The model gc1 represents the case where the $B$-field is dominated equally by the azimuthal and radial components.
Since the model gc3 represents a limiting case in which $B_{\rm z}$ almost dominates near the disk mid-plane at $z/r\le0.06$ (but $B_{\rm \phi}$ cannot be ignored), it is possible that the  initial $B_{\rm z}$ from the replenishing inflow is still preserved in the SW-lobe.
We also find that the NE-lobe and the region E can be better fitted by the model gc4.
The model gc4 implies that the azimuthal field ($B_{\phi}$) becomes more important relative to the radial field ($B_{\rm r}$) as the vertical field ($B_{\rm z}$) decreases in magnitude in the NE-lobe and region E. The main support against gravity can then come from the hydrostatic $B$-field pressure in the NE-lobe.
As we have noted in Section 2.2, there is an  apparent sudden jump in $\Phi_{\rm B}$ from the NE-lobe to the southern extension. The different field-line properties demonstrate that the sudden jump in $\Phi_{\rm B}$ between the NE-lobe (region E) and the southern extension is physical and real. Moreover, the NE-lobe and region E may form a continuous streamer.

In summary, although the model gc1 can describe the overall large-scale field, sub-structures associated with filamentary structures show systematic deviations from the overall axisymmetrical field. 
Our analysis shows that the $B$-field in the CND obtained by the 850 $\micron$ polarization reveals more detailed sub-structures than the far-infrared polarization data. These sub-structures are associated with the pc-scale streamers, as indicated by our high-resolution SMA maps, and they are likely at the origin of the systematic deviations form the overall gc1 model. 
The different smaller-scale features
in the $B$-field might  also be related to the accretion history of the past incoming flows, i.e., the age of each of the CND structures.
This possibility echoes with the previous report by \citet{martin12}, who found that the northeast extension (close to NE-lobe) has been mostly ionized in the northern arm of the mini-spiral and less molecular gas is left than in the southwest lobe. The SW-lobe likely will begin to be ionized when it approaches the central region. 
Since a single WK-model cannot explain all detailed features of the observed $B$-field morphologies, those regions that deviate from the model gc1 may also indicate different stages of accretion.

\subsection{$B$-field -- Linking Mini-Spiral and CND?}

\begin{figure*}[t]
\begin{center}
\epsscale{0.5}
\includegraphics[angle=0,scale=0.7]{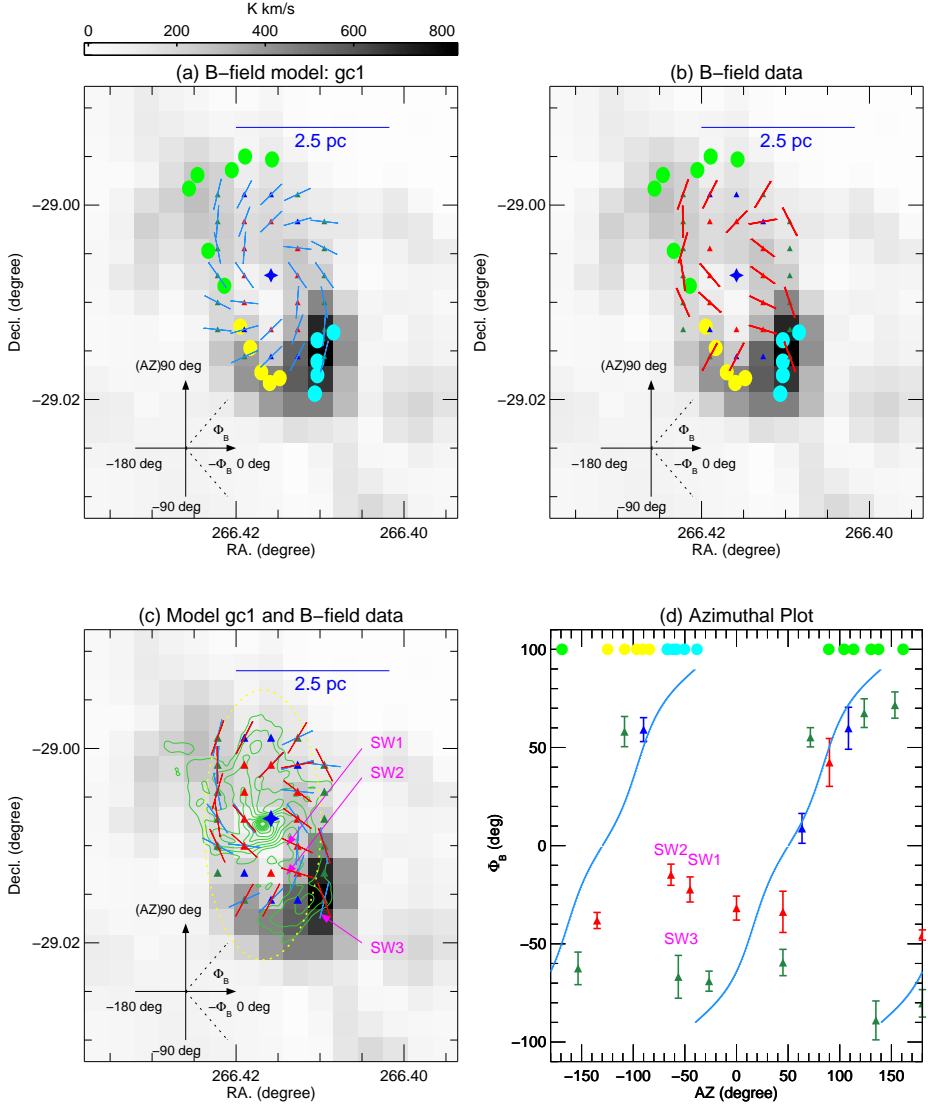}\vspace{-1cm}
\caption[]{\linespread{1}\selectfont{}\small
(a): $B$-field model (gc1) of the interior CND and the mini-spiral shown in blue segments overlaid on the JCMT HCN($J=4-3$) map. The $B$-field is sampled with the same annuli (red, blue, and green triangles). A scale bar of 2.5 pc is shown with the blue line.
(b) The SCUPOL $B$-field data are shown with red segments.
Note that 19 out of 20 segments have the mean uncertainty of $B$-field angle $\sim7.5^{\circ}$ and are $\sigma_{\rm P}\ge3$. The segments labeled with SW1, SW2, and SW3 are above $\sigma_{\rm P}\ge3$.
(c) The $B$-field model (blue segments) and the data (red segments) are overlaid on the HCN($J=4-3$) map. The yellow-dotted ellipse marks the boundary of the mini-spiral in green contours.
(d) Observed $az$-$\Phi_{\rm B}$ correlation of the three annuli (red, blue, and green triangles). Error bars of $\pm1\sigma$ are shown. The $B$-field model (gc1) is overlaid with the blue line.
}
\label{fig-model-data-6cm}
\end{center}
\end{figure*}

\begin{figure*}[ht]
\begin{center}
\epsscale{0.5}
\includegraphics[angle=0,scale=0.5]{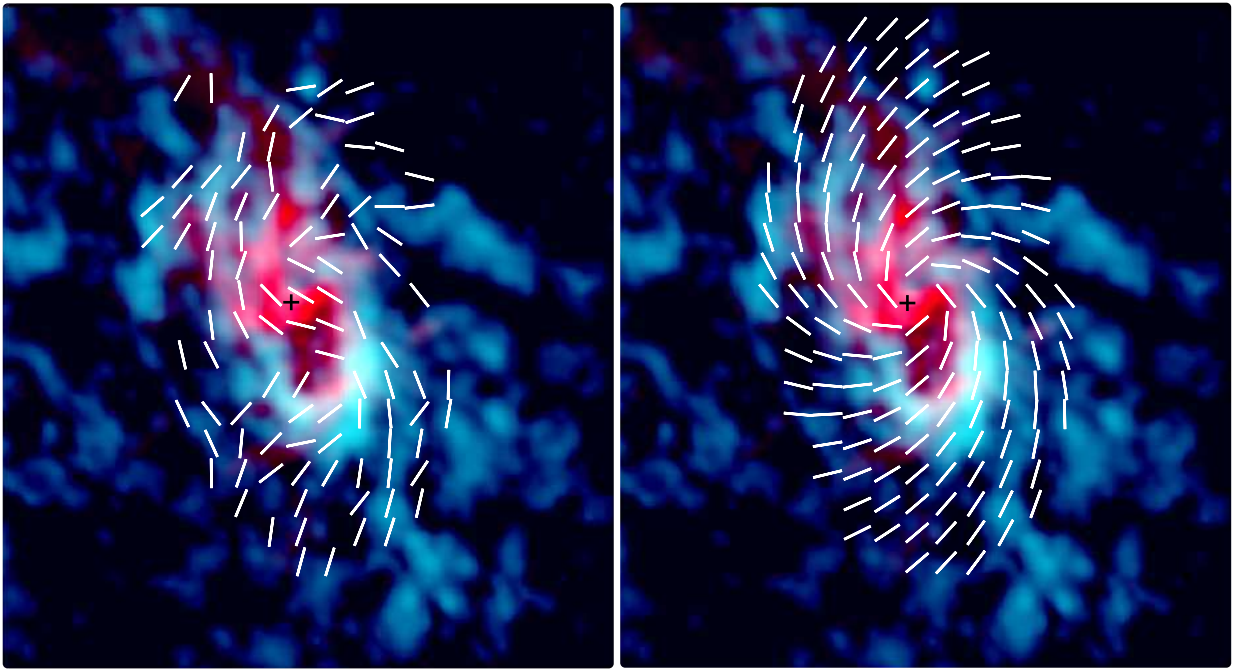}\vspace{0.3cm}
\caption[]{
Color-composite image of SMA HCN($J=4-3$) tracing the CND (blue) and VLA 6 cm
map tracing the mini-spiral (red). The $B$-field of the JCMT data and model gc1 are overlaid with the white segments in the left and right panel, respectively. The location of SgrA* is labeled with the black cross.}
\label{fig-rgb}
\end{center}
\end{figure*}

In Figure~\ref{fig-model-data-6cm}, we have found that the $B$-field lines trace not only the outer neutral streamers towards the CND but also the inner ionized mini-spiral. In the $az$-plot, the smooth connection of the $B$-field from the CND to the inner mini-spiral can be described by the same model gc1. This suggests that the CND and the mini-spiral are connected structures with the same underlying  $B$-field.
In Figure~\ref{fig-rgb} we display the color-composite map of the CND and the mini-spiral with the $B$-field segments overlaid.
In earlier work, the northern arm of the mini-spiral and the NE-lobe of the CND  were already shown to be kinematically connected to the CND 
with closely agreeing velocities \citep{chris}. In agreement with our result, 
the earlier 10 $\micron$ polarization measurements in \citet{aitken86,aitken91} and \citet{roche18} also found that the $B$-field within the northern arm of Sgr A West is parallel to it. 

Our comparison of model and data shows that the CND and the mini-spiral might be linked by field lines. However, two data points (SW1, SW2 in panels c and d in Figure~\ref{fig-model-data-6cm}) show large deviations from the model gc1. We note that the $\Phi_{\rm B}$ of SW1 and SW2 are 
significantly different from the earlier published 850 $\micron$ polarization data in \citet{aitken00}, which show 
a clearer north-south orientation in agreement with model gc1.
There are several comments that can be made here:
(1) The small $\Phi_{\rm B}$ of SW1 and SW2 as presented in our figures might indeed indicate that the vertical field component $B_{\rm z}$ can be important here as suggested by gc5; (2) the  SW-lobe (SW3) in the CND also has a large deviation around 40$\degr$ from the gc1 model. Since the SW-lobe may be undergoing gas replenishment \citep{hsieh17}, the SW1, SW2, and SW3 all together might be contaminated by superimposed components of the incoming streamer; (3) calibration uncertainties might explain the differences between \citet{aitken00} or \citet{matthews09} (note that \citet{matthews09} adopted larger grid than \citet{aitken00}).
Although we cannot further elucidate on these comments with only these few data points and the current limited  resolution and sensitivity, 
we can still conclude that 
the $B$-field morphology of the mini-spiral is overall in agreement with a continuation of the $B$-field morphology 
from the CND. 
Hence, the $B$-field might have the potential to guide the motions of the ionized blobs/filaments originated from the CND and to form the mini-spiral. Moreover, the CND and the mini-spiral could possibly even be understood together as a ``disk'' instead of a ``ring'', with an exterior molecular phase and an interior ionized phase.

\subsection{How Important is the $B$-Field in CND and Mini-Spiral?}

One way to quantify the dynamical importance of the $B$-field is the plasma parameter, $\beta_{\rm Plasma}$, 
that measures the ratio between thermal and magnetic pressure and can be expressed as

\begin{equation}
\beta_{\rm Plasma}\equiv\frac{c_{\rm s}^2}{v_{\rm A}^2}=\frac{4\pi P_{\rm th}}{B^2},
\end{equation}
where $c_{\rm s}$ and $v_{\rm A}^{2}=B^{2}/(4\pi\rho_{\rm i})$ are the sound speed and  the Alfv$\acute{\rm e}$n velocity, respectively. $\rho_{\rm i}$ is the total mass density of the charged particles. $P_{\rm th}$ is the mean thermal pressure of the ionized gas. Taking a $B$-field strength of $B=$1 mG measured from the Zeeman observation in \citet{plante94} and a value  for the thermal pressure $P_{\rm th}/k\sim4\times10^{8}$ cm$^{-3}$ K determined in \citet{hsieh16} \citep[based on an electron density $n_{\rm e}=7.8\times10^{4}$ cm$^{-3}$ and an electron temperature $T_{\rm e}=5000$ K which are in the ranges derived in][]{zhao10},  we find $\beta_{\rm Plasma}\sim 0.7$.  This indicates that a milli-Gauss magnetic field can be important for the dynamical evolution of the gas inflow in the GC.

It has to be noted that the above estimate, strictly speaking, is based on a single localized assessment of the role of the $B$-field because the measurement of the $B$-field strength is from a single-pointing Zeeman observation 
\citep{plante94} that is covering
a limited small region in the GC.  Hence, this estimate does not take into account in any way the morphological features 
in the $B$-field that we have presented and analyzed in the earlier sections. 
Our goal here is to additionally utilize the overall
B-field morphology to assess the dynamical role of the $B$-field. 
Making use of the azimuthal-to-vertical field strength ratio $\alpha$  
($=4\pi\rho_{\rm i} v_{\phi} v_{\rm r}$/$B^{2}$) and 
the inflow-to-rotational velocity ratio $\epsilon$ ($=v_{\rm r}$/$v_{\phi}$) from the WK-model, 
the plasma parameter $\beta_{\rm Plasma}$ can be rewritten as

\begin{equation}
\beta_{\rm Plasma}\equiv\frac{c_{\rm s}^2}{v_{\rm A}^2}=\frac{\alpha}{\epsilon}\frac{c_{\rm s}^2}{v_{\phi}^2}.
\end{equation}

For our overall best-fit model gc1, $\alpha=40$ and $\epsilon=0.25$.
In \citet{hsieh17} we observationally determined an azimuthal velocity $v_{\phi}=110$ km s$^{-1}$.
\citet{great} estimated the gas temperature in the CND to be from 175 K to 325 K.
The corresponding $c_{\rm s}$ then ranges from 1.3 to 1.8 km s$^{-1}$. 
The $\beta_{\rm Plasma}$ obtained with this thermal sound speed is $\sim0.02-0.05$ for model gc1. This number seems to be too small because if $\beta_{\rm Plasma}\le0.1$, the $B$-field would behave like an almost rigid wire. In reality, the CND is additionally subject to turbulent motions. We estimate the linewidth of the cloud in the CND with the velocity linewidth-size relation determined by \citet{miyazaki00}. In the marginally resolved JCMT HCN($J=4-3$) map, the largest cloud size is $\sim1$ pc (radius). The corresponding linewidth is $\sim10$ km s$^{-1}$ based on their empirical relation. A $c_{\rm s}$ $\sim$10 km s$^{-1}$ yields $\beta_{\rm Plasma}\sim1.3$. If the linewidth-size relation determined by \citet{miyazaki00} holds in the CND, adopting a 1 pc cloud size with a  10 km s$^{-1}$ line widths might be an upper limit because most clumps seen in the SMA map are smaller.  Therefore, smaller line widths in e.g., compact clouds would lead to an even smaller $\beta_{\rm Plasma}$, and hence might hint at an even more dominating role of the $B$-field in compact clouds.
Besides the local estimate based on a Zeeman field strength measurement, we can thus conclude that also globally towards and across the CND and onwards to the mini-spiral the B-field appears to be dynamically significant when compared to thermal and turbulent pressure.

Lastly, a comparison between  gravity and $B$-field in the WK-model can be elucidated by the ratio of $v_{\phi}$ and $v_{\rm A}$. The best-fit model gc1 has $\alpha=40$ and $\epsilon=0.25$, leading to $v_{\phi}$/$v_{\rm A}=(\alpha/\epsilon)^{0.5}\sim13$. This indicates that the rotation driven by the enclosed mass is about 10 times faster than the Alfv$\acute{\rm e}$n velocity, suggesting that the CND is a rotation-dominated disk and the $B$-field is less important in affecting the disk rotation.

\section{Summary}

We utilize the JCMT SCUPOL 850 $\micron$ polarization data to investigate the $B$-field structure of the CND in the GC. 
The SCUPOL 850 $\micron$ data show a highly improved and broader polarization coverage at a higher resolution than earlier 100 $\micron$. The 850 $\micron$ data also provide the highest resolution (20$\arcsec$) submillimeter polarimetry data that are probing the B-field morphology  towards the CND, its streamers, and the inner mini-spiral.
On a large scale, the $B$-field shows a clear predominantly azimuthal morphology. In a detailed look, the 850 $\micron$ polarization reveals filamentary-like structures which are not revealed in the  100 $\micron$ and 350 $\micron$ polarization data. These sub-structures in the $B$-field morphology are associated with the earlier identified streamers in the CND (the SW-lobe, the southern extension, and the NE-lobe). These streamers were previously shown to consist of distinct Keplerian components in the SMA CN map.
Moreover, the innermost observed $B$-field structure also appears to trace and align with the ionized mini-spiral located inside the CND.  This suggests that there might be one underlying $B$-field structure that is connecting the CND with its streamers and the inner mini-spiral as shown in Figure~\ref{fig-rgb}.

Overall on a large scale, the observed $B$-field morphology is well described by a self-similar axisymmetric disk model where the radial infall velocity is one quarter of the rotational velocity (WK model gc1). However, a detailed comparison of the deviation from the model gc1 and individual streamers shows that the model gc3 and gc4 can describe the SW-lobe and the NE-lobe of the CND better than model gc1. Our interpretation is that in the SW-lobe, the vertical $B_{\rm z}$ 
component is still preserved along the incoming streamer. In the NE-lobe, the $B$-field is dominated by the azimuthal field component $B_{\phi}$. The variation of the field configurations among these streamers might suggest an accretion history or different stages of accretion in the CND. The NE-lobe has been mostly ionized in the northern arm of the mini-spiral and less molecular gas is left than in the SW-lobe. On the other hand, the SW-lobe likely will begin to be ionized when it approaches the central region.

We also estimate the dynamical importance of the $B$-field with the plasma parameter ($\beta_{\rm Plasma}$).  Based on a single-pointing Zeeman 
field strength measurement, we derive $\beta_{\rm Plasma}\sim0.7$. 
Based on the global $B$-field morphology that constrains the azimuthal-to-vertical field strength ratio ($\left|B_{\phi}/B_{\rm z}\right|$) to be around 40 combined with a measurement of the azimuthal velocity ($v_{\phi}$), we estimate $\beta_{\rm Plasma}\sim1.3$ as an upper limit. $\beta_{\rm Plasma}$ being of order unity indicates that the $B$-field appears to be dynamically significant towards the CND and also onwards to the inner mini-spiral as compared to thermal/turbulent pressure. Although the CND is a rotation-dominated disk, the coherent underlying field geometry and $\beta_{\rm Plasma}\lesssim1$ suggest that the $B$-field can help to guide the motions of the ionized blobs/filaments originated from the CND, possibly form the mini-spiral, and provide substantial support to structures as compared to thermal and turbulent pressure. 

We emphasize  that the $B$-field morphology traced at 850 $\micron$ shows dramatically more sub-structures than at 100 $\micron$ and 350 $\micron$.
The observations at shorter wavelengths possibly trace warmer dust components, while the 850 $\micron$ emission is dominated by a cooler and likely denser dust component, i.e., the CND and the inflow material. Our results reveal the uniqueness of using 850 $\micron$ polarization to trace cooler disk components / streamers, and also show that the $B$-field can play an important role for the gas flows and accretion in the GC. 

\acknowledgements
We thank the anonymous referee for comments  which have helped  to improve the manuscript. 
 P.-Y. H. is supported by the Ministry of Science and Technology (MoST) of Taiwan through the grants MoST 105-2811-M-001-141,  MoST 106-2811-M-001-136.  P. M. K. acknowledges support from MoST 104-2119-M-001-019-MY3 and from an Academia Sinica Career Development Award. W.-T.K. was supported by the National Research Foundation of Korea (NRF) grant funded by the Korea government (MEST; No. 3348-20160021).


\clearpage
\appendix

\section{Comparison with Previous Data}

\begin{figure*}[ht]
\begin{center}
\includegraphics[angle=0,scale=0.9]{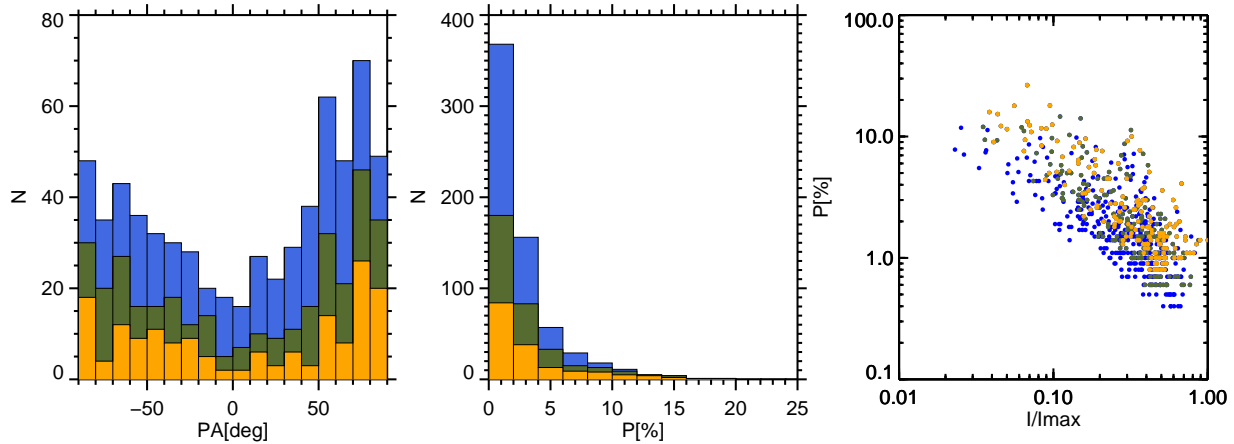}
\caption[]{Left: distributions of $B$-field position angles (PA) of SCUPOL 850 $\micron$ data.  
Middle: distributions of polarization percentages (P).  
Right: Stokes $I$, normalized to $I_{\rm max}$, versus polarization percentage $P$. In all panels, blue, olive, and orange present the data above $\sigma_{\rm P}\ge$ 2, 3, and 4, respectively.}
\label{fig-histogram}
\end{center}
\end{figure*}

\begin{figure*}[ht]
\begin{center}
\includegraphics[angle=90,scale=0.7]{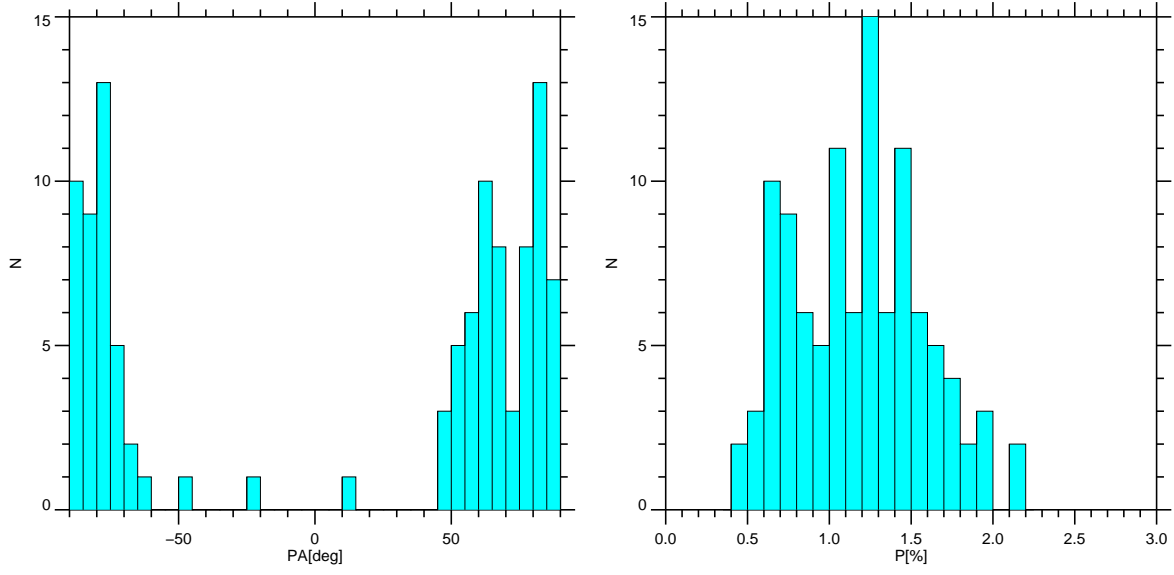}
\caption[]{Left: distribution of the position angles (PA) of the $B$-field of the HERTZ 350 $\micron$ polarization data in the GC  \citep{novak00}. Only data above $\sigma_{\rm P}\ge$ 3 are presented in their paper. Right: distribution of polarization percentages.}
\label{fig-histogram-hertz}
\end{center}
\end{figure*}

In Figure~\ref{fig-histogram} we show the 850 $\micron$ SCUPOL histograms of the polarization percentages ($P$) and position angles ($PA$) of the $B$-field ($\Phi_{\rm B}$) with $p/dp~(=\sigma_{\rm P}) \ge2, 3, 4$ as presented in the three panels in Figure~\ref{fig-frans}. The overall distributions do not vary significantly from $\sigma_{\rm P}$ of 2 to 4 cuts.
In particular, this indicates that overall structural features in the $B$-field, measured  by position angles PA, are preserved from 4$\sigma_{\rm P}$ to 2 $\sigma_{\rm P}$, and the different cuts in $\sigma_{\rm P}$ are unlikely mimicking or  introducing  wrong global large-scale trends in the B-field.  We note this finding in order to make best use of the available data. 
Therefore, in order to work with the maximum number of data points, we will present the $B$-field segments using $\sigma_{\rm P} \ge2\sigma$ in this paper. The majority of the $P$ is around $1.5\%$ with a tail extending to $10-25\%$. In the right panel in Figure~\ref{fig-histogram}, the largest $P$ are found in the faintest Stokes I regions.
This indicates that the generally observed $P-I$ anti-correlation in star-forming regions  \citep[see, e.g., discussion in][]{tang13} also holds in the GC environment. Hence, similar dust grain properties and alignment with the magnetic field can probably be expected in the GC.
Histograms of the 350 $\micron$ polarization data  \citep{novak00} are shown in Figure~\ref{fig-histogram-hertz}. 
Comparing to the 850 $\micron$ data, we find that the 350 $\micron$ data indicate mostly a north-south orientation
of the $B$-field with a $PA$ distribution peaking around $\pm90^{\circ}$. While this prevailing north-south orientation
is still also apparent in the 850 $\micron$ $PA$ distribution, the 850 $\micron$ data additionally show finer 
sub-structures which are central for our discussion in this work.

\begin{figure*}[ht]
\begin{center}
\epsscale{0.5}
\includegraphics[angle=0,scale=0.4]{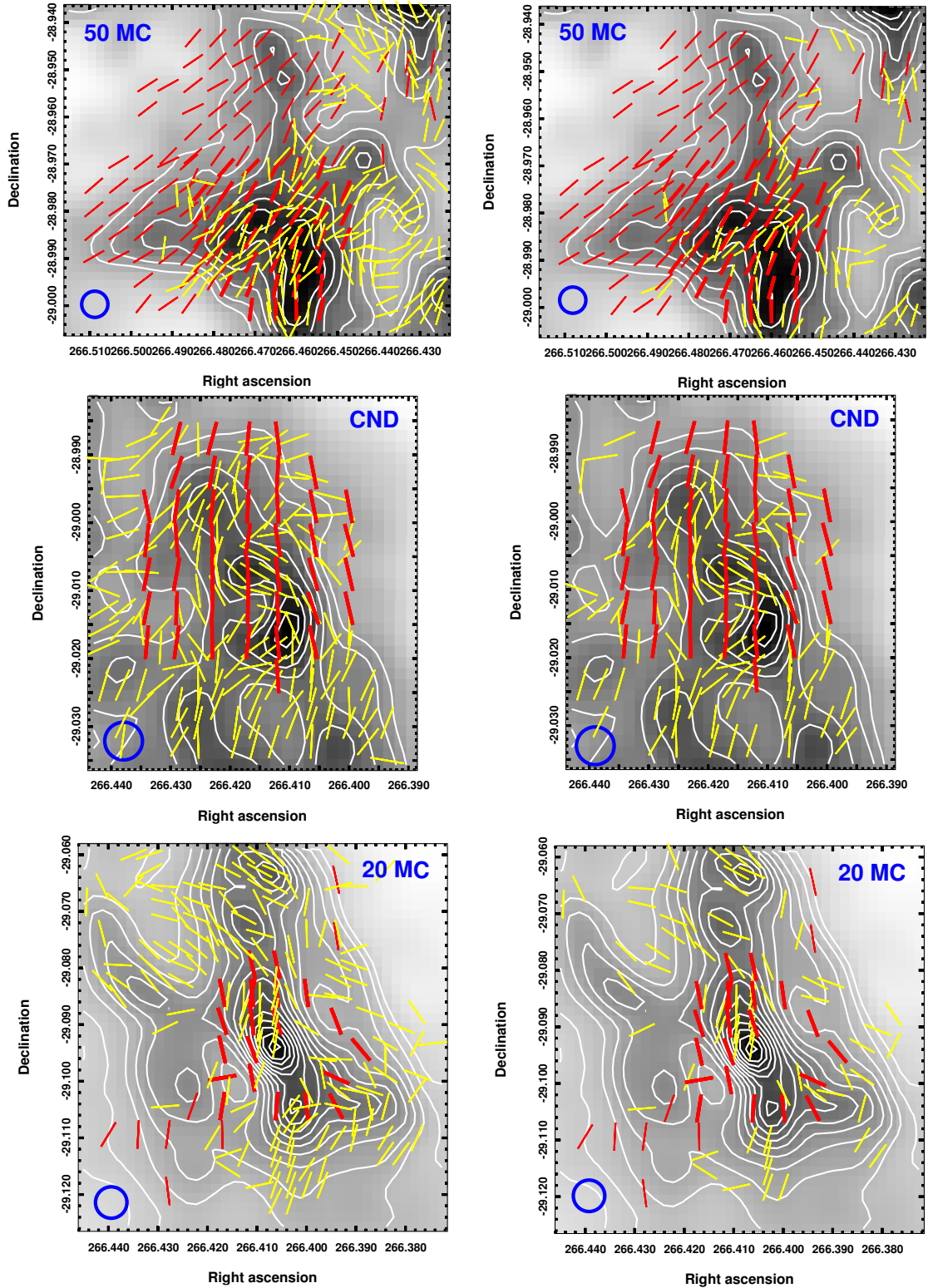}\vspace{0.5cm}
\caption[]{Comparison of the $B$-field probed by 850 $\micron$ SCUPOL data (yellow segments) and 350 $\micron$ HERTZ data (red segments; sampling with 20$\arcsec$ grid) towards the 20/50 MC and the CND. The contour levels are the same as Figure~\ref{fig-frans}. Left and right columns show the JCMT data with $\sigma_{\rm P}\ge$ 2 and 3, respectively. The thin and thick red segments present the data from \citet{novak00} and \citet{chuss03}, respectively. The 20$\arcsec$ HERTZ and SCUPOL beam is displayed with a blue circle.}
\label{fig-compare}
\end{center}
\end{figure*}

Figure~\ref{fig-compare} compares the $B$-field from the 850 $\micron$ SCUPOL and the 350 $\micron$ HERTZ data \citep[][beam=20$\arcsec$]{novak00,chuss03} toward the 20/50 MC and the CND.
The orientation of the $B$-field in the 20 MC is mostly north-south, both at 850 $\micron$ and at 350 $\micron$. For the 50 MC and the CND, \citet{chuss03} mention that the $B$-field in these two regions shows a coherent $X$-shaped polarization feature at 350 $\micron$ which traces the arched filaments in the GC. The $B$-field morphology in the 50 MC at 850 $\micron$ is similar to that at 350 $\micron$.
However, the difference of the $B$-field morphology in the CND between the 350 $\micron$ and the 850 $\micron$ data may suggest different origins for the polarized signals. In the CND region, \citet{novak00} estimated that one third of the 350 $\micron$ emission is not from the CND. The $B$-field at 850 $\micron$ seems to trace the inner sub-structures shown in the continuum map, while the $B$-field at 350 $\micron$ likely traces more large-scale structures associated with the arched filaments as suggested by \citet{chuss03}. 
Hence, the 350 $\micron$ data likely trace warmer and possibly upper-layer components of the CND, while the 
850 $\micron$ data likely trace the inner cooler sub-structures (e.g., streamers) of the CND.


\clearpage

\section{Predicted $B$-field Morphologies with the WK-Model}

\begin{figure*}[ht]
\begin{center}
\epsscale{0.5}
\includegraphics[angle=0,scale=0.8]{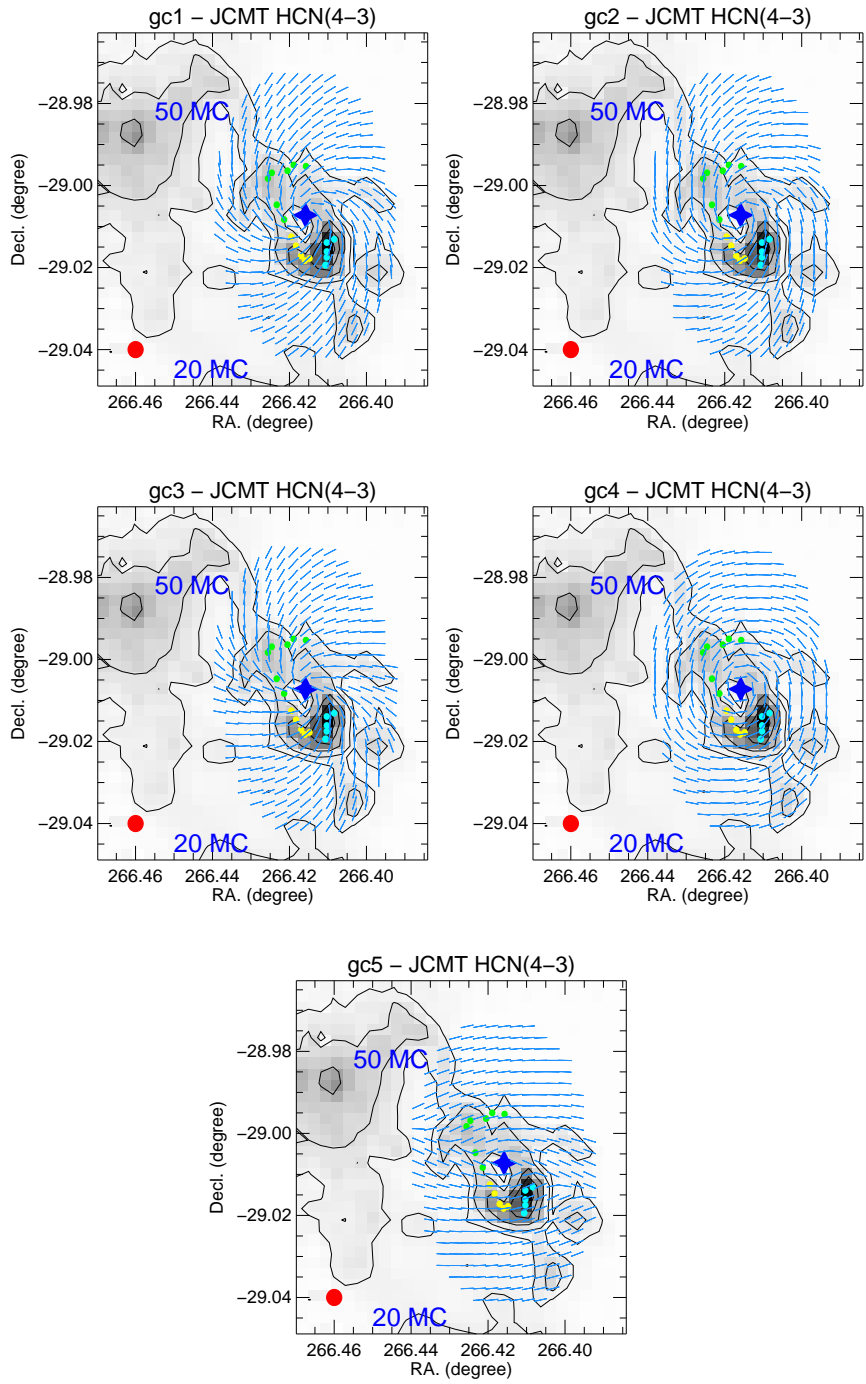}\vspace{-0.5cm}
\caption[]{\linespread{1}\selectfont{}\small The WK-models (blue segments; gc1, gc2, gc3, gc4, gc5) overlaid on the JCMT HCN($J=4-3$) map. The beam size of 14$\arcsec$ is displayed with the red circle. Previously identified features associated with the CND -- northeast lobe (NE-lobe), southwest lobe (SW-lobe), and southern extension are indicated with the green, cyan, and yellow dots, respectively. SgrA* is marked with the blue star.
}
\label{fig-4model}
\end{center}
\end{figure*}

\begin{figure*}[ht]
\begin{center}
\epsscale{0.5}
\includegraphics[angle=0,scale=0.95]{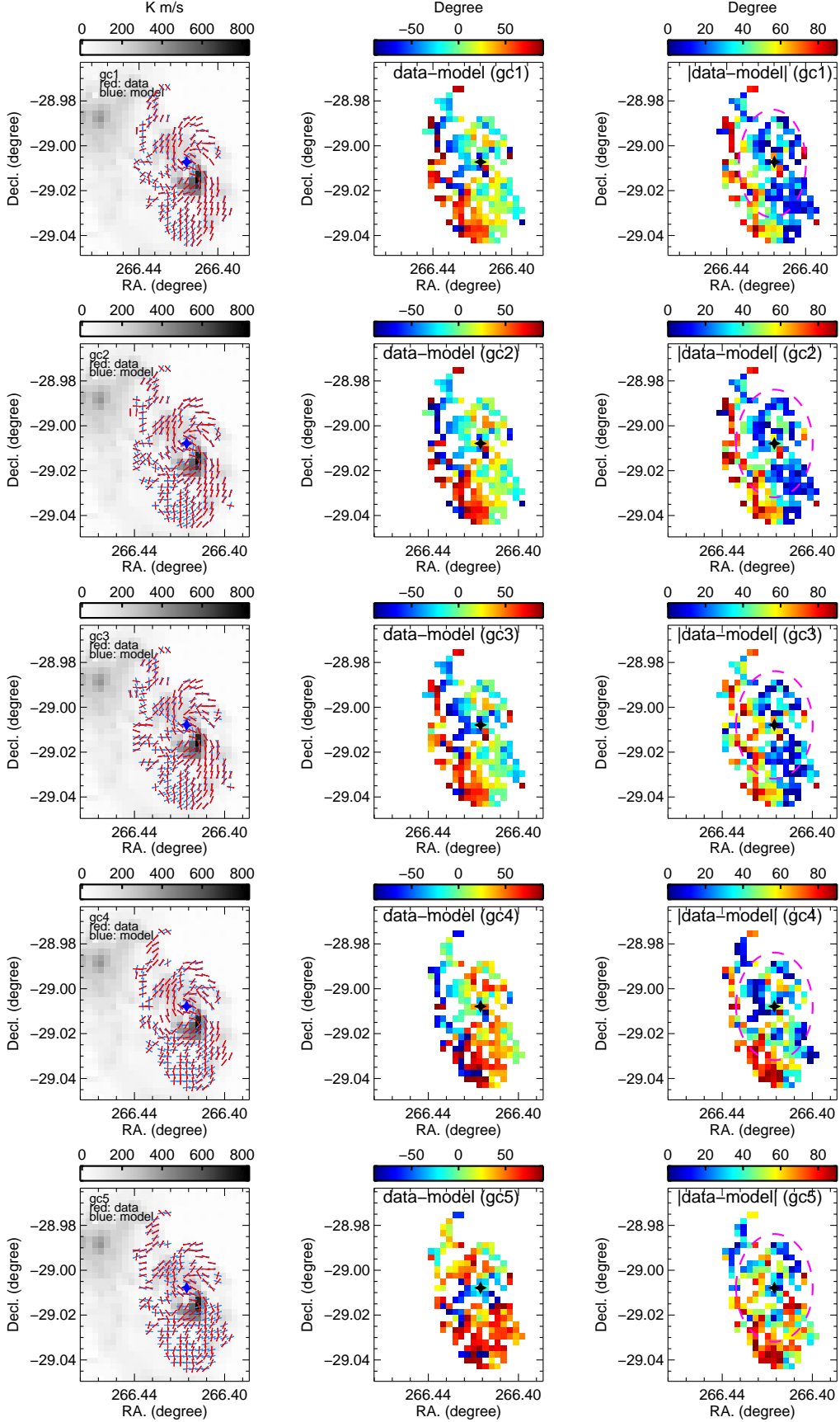}
\caption[]{
Left: Observed $B$-field (red segments) and WK-models (blue segments) overlaid on the JCMT HCN($J=4-3$) map. The models gc1 to gc5 are shown from top to bottom.
Middle: Differences between observed $B$-field and model $B$-field ($\phi_{\rm B}$ - $\phi_{\rm model}$) are shown in color.
Right: Same as in the middle panel, but for absolute differences.
}
\label{fig-model-gc1234}
\end{center}
\end{figure*}

\begin{figure*}[ht]
\begin{center}
\epsscale{0.5}
\includegraphics[angle=0,scale=0.7]{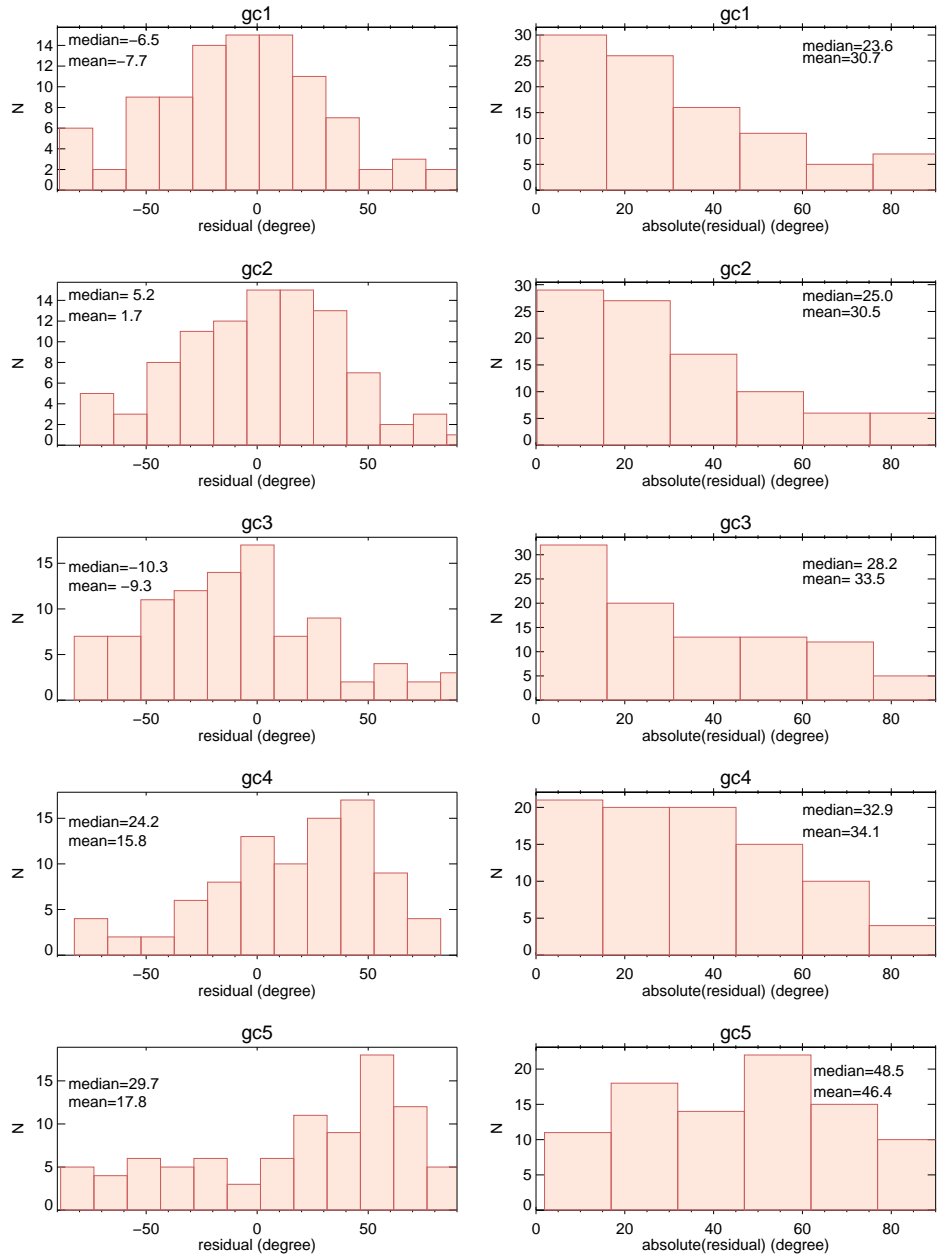}
\caption[]{Histograms of the differences between models and data shown in Figure~\ref{fig-model-gc1234}. A region within 80$\arcsec$ centered on the CND is used for the calculation of mean and median values to avoid the possible confusion from the 20 MC. The mean and median values for gc1 to gc5 are shown from top to bottom. Left and right columns show the real and absolute values, respectively.
}
\label{fig-model-hist}
\end{center}
\end{figure*}

\begin{figure*}[ht]
\begin{center}
\epsscale{0.5}
\includegraphics[angle=0,scale=0.7]{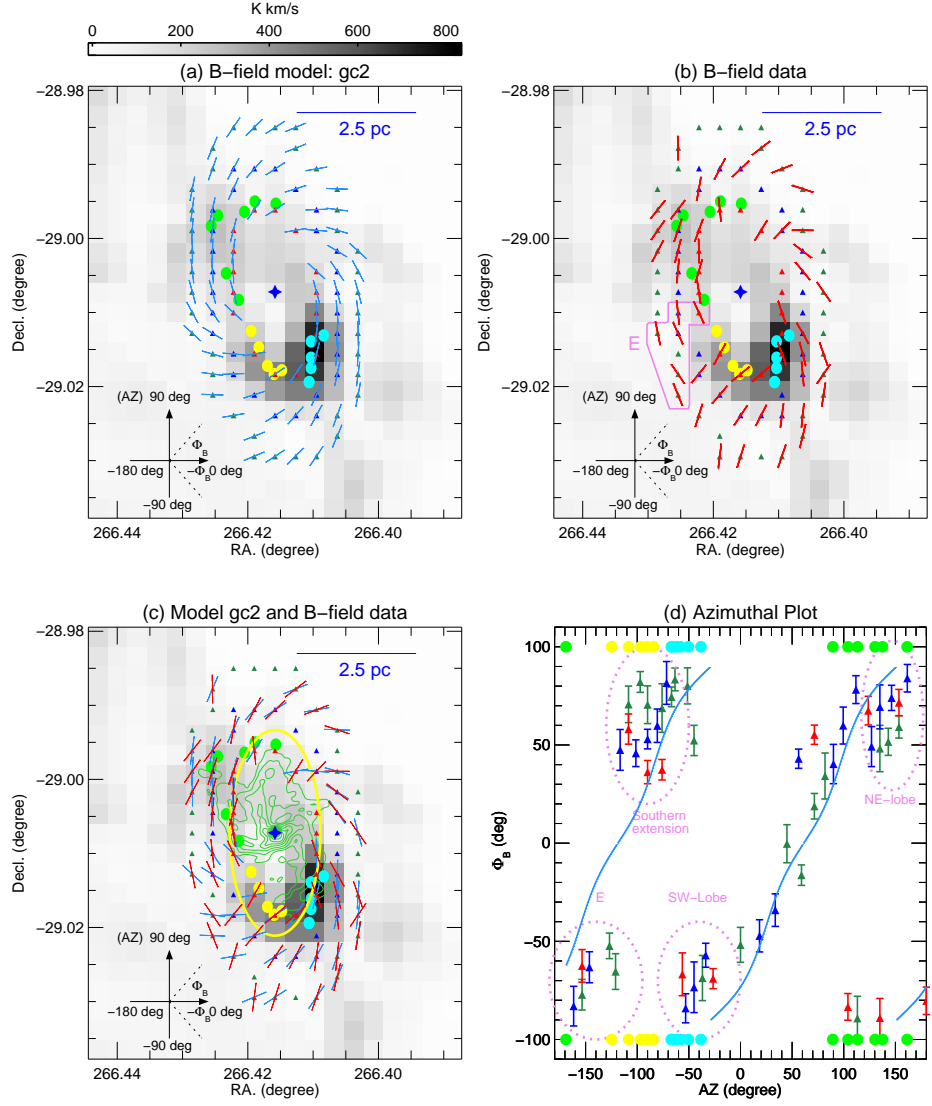}\vspace{-1cm}
\caption[]{(a): $B$-field model (gc2) in blue segments overlaid on the JCMT HCN($J=4-3$) map. The $B$-field is sampled with the same annuli (red, blue, and green triangles) as shown in Figure~\ref{fig-bf-vg}. A scale bar of 2.5 pc is shown with the blue line.
(b) The SCUPOL $B$-field data are shown with red segments.
(c) $B$-field model (blue segments) and data (red segments) overlaid on the HCN($J=4-3$) map. The yellow ellipse marks the boundary of the mini-spiral in green contours. 
(d) $az$-$\Phi_{\rm B}$ correlation of the three annuli (red, blue, and green triangles). Error bars of $\pm1\sigma$ are shown. The $B$-field model (gc2) is overlaid with the blue line.
}
\label{fig-model-az-gc2}
\end{center}
\end{figure*}

\begin{figure*}[ht]
\begin{center}
\epsscale{0.5}
\includegraphics[angle=0,scale=0.7]{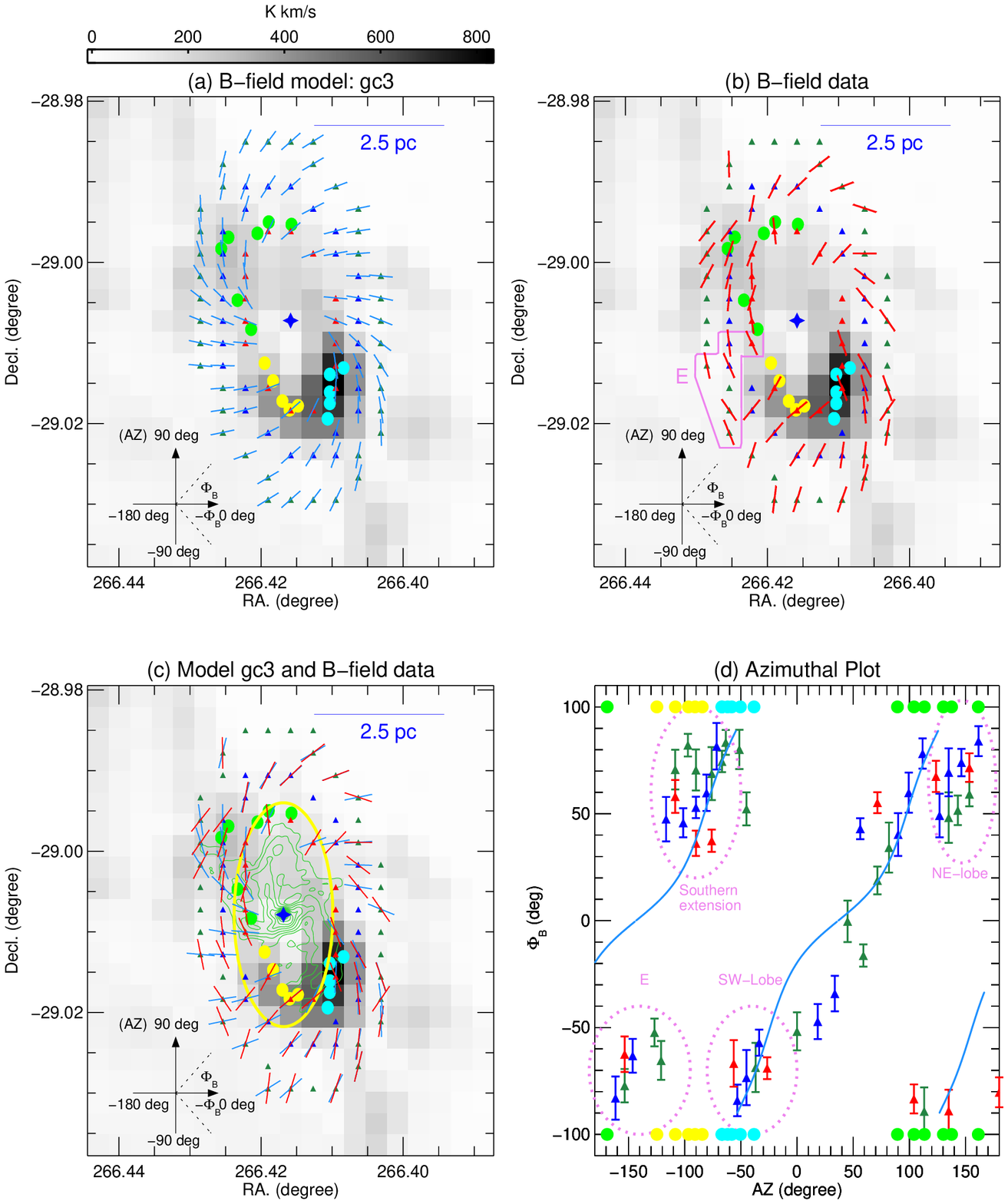}\vspace{-1cm}
\caption[]{Identical to Figure~\ref{fig-model-az-gc2} but for $B$-field model gc3.
}
\label{fig-model-az-gc3}
\end{center}
\end{figure*}

\begin{figure*}[ht]
\begin{center}
\epsscale{0.5}
\includegraphics[angle=0,scale=0.7]{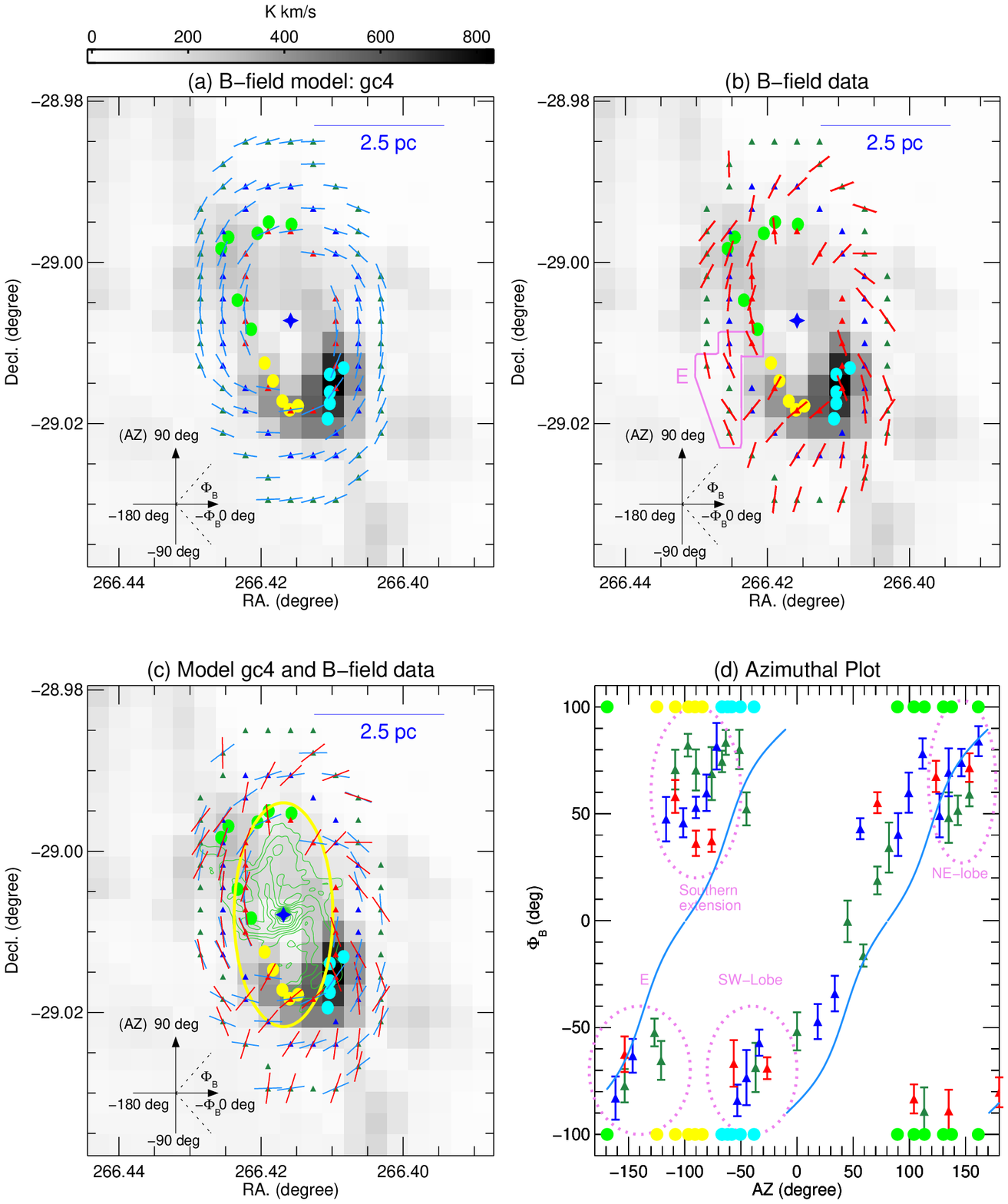}\vspace{-1cm}
\caption[]{Identical to Figure~\ref{fig-model-az-gc2} but for $B$-field model gc4.
}
\label{fig-model-az-gc4}
\end{center}
\end{figure*}

\begin{figure*}[ht]
\begin{center}
\epsscale{0.5}
\includegraphics[angle=0,scale=0.7]{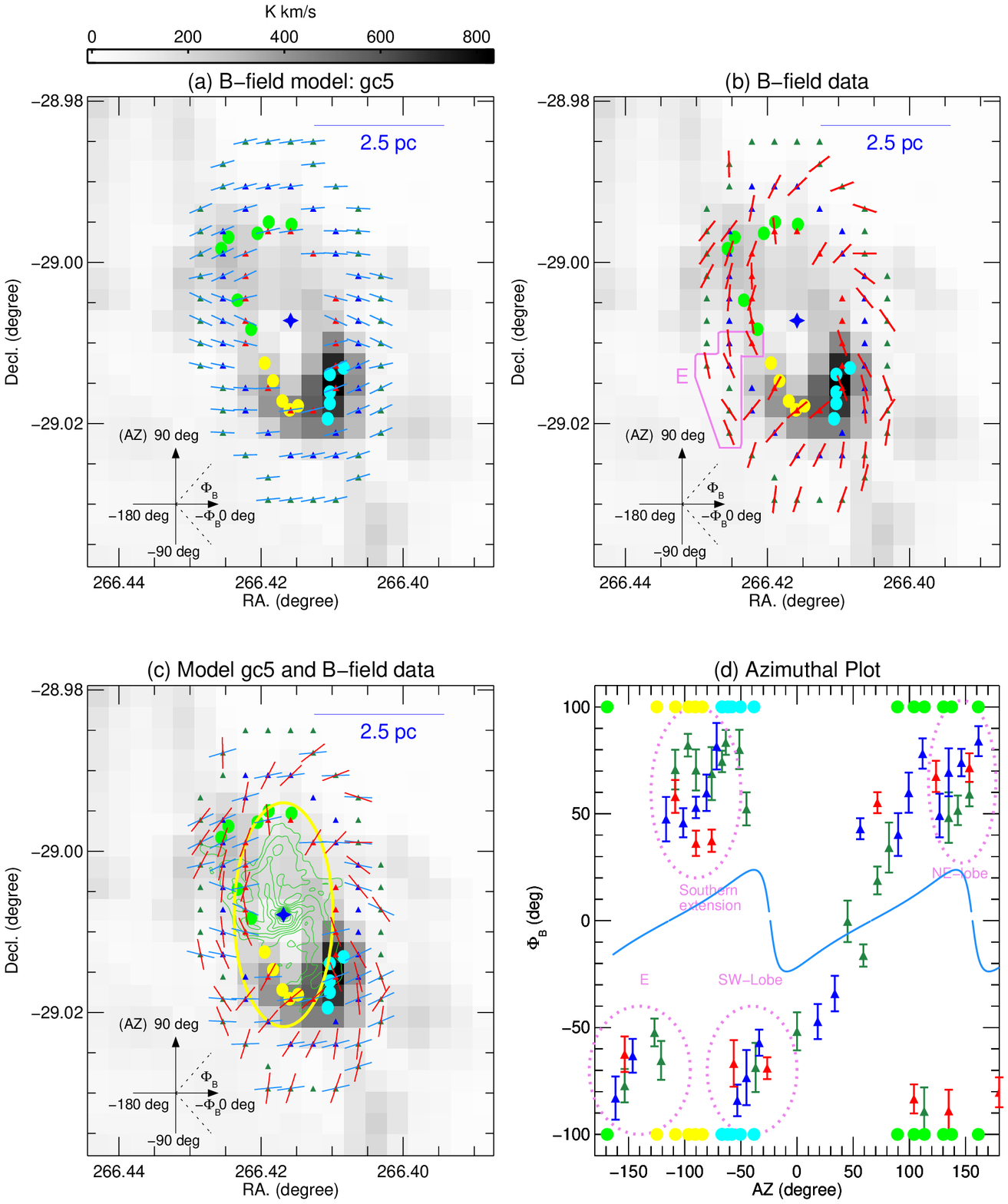}
\caption[]{Identical to Figure~\ref{fig-model-az-gc2} but for $B$-field model gc5.
}
\label{fig-model-az-gc5}
\end{center}
\end{figure*}

The 5 WK-models (gc1 to gc5) \citep[see descriptions in Table 1 in][]{wardle90} are shown in Figure~\ref{fig-4model}. The model $B$-field lines are sampled on a 10$\arcsec$ grid. Comparisons of the models and the data are shown in Figure~\ref{fig-model-gc1234} and Figure~\ref{fig-model-hist}. As in Figure~\ref{fig-model-az-gc1}, Figure~\ref{fig-model-az-gc2} to Figure~\ref{fig-model-az-gc5} show the $az$-plots of the comparisons between the models (gc2 to gc5) and the data. The difference of the orientations between the data and the models, $\Phi_{\rm B}$ - $\Phi_{\rm model}$, is shown in the middle and right panels of Figure~\ref{fig-model-gc1234}. In the CND, the difference is within $\sim20\degr$ for the models gc1, gc2, and gc3. The difference is larger for model gc4 ($\sim40\degr$)  and gc5 ($\sim60\degr$). In all the models, the 20/50 MC show differences larger than $60\degr$, which likely is due to contaminating polarized emission from the 20/50 MC away from the CND, and a likely intrinsically different B-field structure in the 20/50 MC that substantially differs from the axisymmetrical B-field in the CND. 
In Figure~\ref{fig-model-hist}, histograms of the differences of real and absolute values of these 5 models are displayed. The median and mean values are calculated. The difference is toward positive or negative offsets from gc3 to gc5 as compared to the gc1 and gc2. We also run a Chi-square test ($\chi^{2}$/$dof$) for each model, where $dof$ is the degree of freedom. To avoid the regions that might be contaminated by the 20/50 MC, we only compute $\chi^{2}$/$dof$ within $61\arcsec\times85\arcsec$ (magenta ellipse in Figure~\ref{fig-model-gc1234}). The computed $\chi^{2}$/$dof$ for models  gc1 to gc5 are 45.39 ($dof=91$), 115.14 ($dof=91$), 72.60 ($dof=91$), 105.72 ($dof=91$), and 263.71 ($dof=89$), respectively. The model gc1 appears to be the best fit with the smallest $\chi^{2}$/$dof$ ($P\ge99.5\%$).

We note that 
\citet{hilde90} suggested gc1 to be the best-fit configuration for their six polarization segments detected at 100 $\micron$. Later, with about three times more data points, \citet{hilde93} favored model gc2. However, in terms of the mean values (Figure~\ref{fig-model-hist}), gc1 and gc2 are indistinguishable in our analysis based on 850 $\micron$ data. In a more detailed analysis, we find that model gc3 can fit the SW-lobe better than gc1.
Moreover, we also find that the region NE-lobe (middle and inner annuli) is dominated by $B_{\phi}$ and can best be described by model gc4 (see also discussions in Section 4).

\end{document}